\documentstyle[12pt]{article}
\def\ie{{\em i.e.}}
\def\eg{{\em e.g.}}

\def\S{S\hskip-8pt/\hskip2pt}
\def\H{H\hskip-10pt/\hskip2pt}

\def\beq{\begin{equation}}
\def\eeq{\end{equation}}

\catcode`\@=11 

\def\VEV#1{\left\langle #1\right\rangle}
\def\vev#1{\left\langle #1\right\rangle}
\def\lsim{\mathrel{\mathpalette\@versim<}}
\def\gsim{\mathrel{\mathpalette\@versim>}}
\def\@versim#1#2{\vcenter{\offinterlineskip
    \ialign{$\m@th#1\hfil##\hfil$\crcr#2\crcr\sim\crcr } }}
\def\etal{{\em et. al.}}
\def\JL{J. L. Lopez}
\def\DVN{D. V. Nanopoulos}

\def\t1{{\tilde 1}}

\def\cm{\,{\rm cm}}

\def\GeV{\,{\rm GeV}}

\def\lra{\leftrightarrow}
\def\to{\rightarrow}

\def\llra{$\longleftrightarrow$}

\def\PLB#1#2#3{Phys. Lett. B {\bf#1} (19#2) #3}

\def\PRD#1#2#3{Phys. Rev. D {\bf#1} (19#2) #3}

\def\PRT#1#2#3{Phys. Rep. {\bf#1} (19#2) #3}

\def\TAMU#1{Texas A \& M University preprint CTP-TAMU-#1}

\begin{document}
\large
\begin{flushright}
\baselineskip=12pt
{CERN-TH.7260/94}\\
{CTP-TAMU-20/94}\\
{ACT-07/94}\\
\end{flushright}

\begin{center}
{\Huge\bf As time goes by ...\\}
\vglue 0.3cm
{D. V. NANOPOULOS\\}
\vglue 0.2cm
{\em Center for Theoretical Physics, Department of Physics\\
Texas A\&M University, College Station, TX 77843--4242, USA;\\}
{\em Astroparticle Physics Group, Houston Advanced Research Center
(HARC), The Mitchell Campus, The Woodlands, TX 77381, USA;\\}
{\em CERN Theory Division, 1211 Geneva 23, Switzerland\\}
\baselineskip=12pt

\vglue 0.2cm
{ABSTRACT}
\end{center}
{\normalsize
A rather simple and non-technical exposition of our new approach to {\em
Time, Quantum Physics, Black-Hole dynamics}, and {\em Cosmology}, based on
non-critical string theory, is provided. A new fundamental principle, the
{\em Procrustean Principle}, that catches the essence of our approach is
postulated: the low-energy world is {\em unavoidably} an ``open" system due
to the spontaneous truncation of the {\em delocalized, topological} string
modes in continuous interaction with the low-lying-{\em localized} string
modes. The origin of space-time, the expansion of the Universe, the
entropy increase and accompanied irreversibility of time, as well as the
collapse of the wavefunction are all very neatly tied together. Possible
observable consequences include: quantum relaxation with time of
the Universal, fundamental constants, like the velocity of light $c$ and the
Planck constant $\hbar$ decreasing towards their asymptotic values, and the
cosmological constant $\Lambda_C$ diminishing towards zero; possible
violation of {\em CPT} invariance in the $K^0-\bar K^0$ system, possible
apparent non-conservation of angular momentum, and possible
loss of quantum coherence in SQUID-type experiments.}
{\rightskip=3pc
 \leftskip=3pc
\noindent
}
\begin{flushleft}
\baselineskip=12pt
{CERN-TH.7260/94}\\
{CTP-TAMU-20/94}\\
{ACT-07/94}\\
May 1994
\end{flushleft}
\hrule
\smallskip
{\small\baselineskip=10pt Extended version of a talk delivered at the Houston
Advanced Research Center, Technical Seminar Series, at Rockefeller University
(N.Y.), and at the Workshop on ``Frontiers in Quantum Field Theory", June 1994,
ITP, Academia Sinica, Beijing, China.}

\newpage

\vspace*{3cm}
\begin{quotation}
``Physicists consider me an old fool, but I am convinced that the future
development of Physics will depart from the present road".\\
\begin{flushright}
Albert Einstein
\end{flushright}

\vspace{3cm}
``It is my opinion that our present picture of physical reality, particularly
in relation to the nature of {\em time}, is due for a grand shake up -- even
greater, perhaps, than that which has already been provided by present-day
relativity and quantum mechanics".\\
\begin{flushright}
Roger Penrose \\
in {\em The Emperor's New Mind}
\end{flushright}
\end{quotation}
\newpage

\setcounter{page}{1}
\markright{}
\pagestyle{myheadings}
\pagenumbering{roman}
\section*{Prolegomena and Epimythion}
During the last fifty years, high-energy physics has gone through many
dramatic changes, mostly positive, that have led us to an unprecendented
understanding of the subnuclear world. From the discovery of the pion ($\pi$)
to the discovery of the top quark ($t$), and from the Lamb-shift to the
LEP-shift(s), all very nicely accommodated in the {\em Standard Model}, the
crown jewel of present day particle physics. So, the basic question arises:
is that it? Certainly not! The Standard Model is certainly a very fundamental
step in the right direction, but it leaves so much to be desired that its
{\em Supersymmetric Grand Unified Theoretical} extension is, for at least
{\em some} of us, an unavoidable {\em must}. As it is, for at least {\em
some} of us (not necessarily the same {\em some} as above), unavoidable
to replace point-like particles by {\em one-dimensional fundamental extended
objects}, strings, \ie, use {\em String Theory}. The raison d'etre of
string theory is rather well-known: natural, automatic unification of all
interactions in nature, including {\em unavoidably} a consistent Quantum
Theory of Gravity. The infrared limit of string theory seems, to many of us,
the correct framework to resolve the usual conundrums of present day particle
physics: {\em origin} of {\em all} masses, angles, coupling constants,
electroweak breaking, etc. Then, is that it? Unless we are desperate
to confirm Dante's ``{\em O insensata cure dei mortali}", we better refrain
from answering yes to this question. Let me remind you that all our Modern
Physics is based on two fundamental, {\em untouchable} principles: the
{\em quantum} one and the (special) {\em relativity} one. Combined they have
provided the spectacularly successful point-like Relativistic Quantum Field
Theory (QFT) framework, the basis of particle physics. {\em Charmed away} by
the phenomenal successes of relativistic QFT, we run into the danger of
conveniently and tacitly neglecting some of the dark corners of quantum physics
and relativity. The absence of a dynamical mechanism for the {\em collapse of
the wavefunction}, related directly to the ``measurement" problem of quantum
mechanics and to the emergence of the classical world from the quantum one,
the {\em arrow of time} shining by its absence from presently available
microscopic theories, the rather incomplete {\em black-hole physics}, not
only do they prevent us from a crystal-clear picture of physical reality, but
they do sustain a {\em Pirandelloish} mystery of reality.

During the last three years, together with John Ellis and Nick Mavromatos,
we have attacked these problems from the premises of non-critical string
theory. To our surprise, this line of research took us very soon to
completely {\em uncharted waters} and we found ourselves facing a plethora
of technical and physical problems. To our initial disbelief, we found that
the {\em untouchable} principles of quantum mechanics and relativity needed
to be touched! We are not talking here about some higher-order corrections to
the Einstein-Hilbert action, or some extra corrections, to say, the
Schr\"odinger equation. What we are talking about here is to {\em abandon}
altogether the language of the wavefunction, to be replaced by the {\em density
matrix}, as well as, to create dynamically an {\em arrow of time} at the
microscopic level. In other words, not only providing each space point with a
clock, \'a la Einstein, but inventing in addition a dynamical mechanism to
synchronize all of them! All these admittedly seemingly {\em incredulous
allegations} are springing off the very fundamental properties of string
theory. String theory is consistently and dynamically endorsed with infinite
symmetries (in sharp contrast with point-like QFT), which mix the different
mass ``levels", including delocalized topological modes, thus triggering
spontaneous truncation down to the {\em observable} localized low-energy string
modes. This {\em unavoidable} truncation renders the low-energy observable
system effectively ``open", thus inhereting the characteristic statistical
properties: increase of entropy, irreversible time, collapse of the
wavefunction, etc. In a way, ``we can have our cake and eat it too": we start
from a complete, consistent string theory, according to all the rules of
quantum theory, no violations of anything, and dynamically the low-energy {\em
observable} world behaves {\em effectively} as an ``open system".

The reaction to our approach, as it was expected, varies from the very negative
to the very positive! The tactics of {\em suppresio veri, suggestio falsi},
and of {\em ``no vogliamo capire"}, have been employed in abundance, which for
us had a very positive outcome: understand much better our engaged technology
and physical issues!

Since our approach touches upon issues that go beyond string theory or even
high-energy physics, I have decided, after some initial resistance, to comply
with the continuous urgings of several friends, {\em to say it like it is}!
I have tried {\em willingly} and {\em deliberately} to minimize the technical
details to the bare bones, and to sacrify mathematical elegance, clarity, and
rigor for the more familiar, at least to me, physical intuition. I would like
to encourage very strongly the interested reader to have a look at
Refs.~\cite{1,2,3} in which a much more technical review of our approach is
provided and to which the present review will hopefully serve as a useful
companion.

\newpage

{\baselineskip=17pt
\tableofcontents}
\newpage

\pagenumbering{arabic}

\baselineskip=16pt

\section{Remembrance of Things Past}

Towards the end of the last century, physics seemed to be in very good shape.
Classical mechanics, Electromagnetic theory, and Thermodynamics were well
under control and most physicists were taking a nohelic point of view that
everything had been understood. There were a few ``small" problems here and
there, but it was widely believed that the existing ``technology" would
eventually resolve them. The most notable ``small" problems were:
\begin{enumerate}
\item Black-body radiation: ultraviolet catastrophe.
\item Michelson-Morley experiment: no ether.
\item Line atomic spectra: do ``a-toms" have substructure(?).
\end{enumerate}

Let me  remind you of the origin of these problems and their eventual
resolution.\\

\subsection{Black-body Radiation}

According to classical physics, the average energy $\VEV{\epsilon}$ at
temperature $T$ of a harmonic oscillator of frequency $w$ is given by
\beq
\vev{\epsilon}_T={\int_0^\infty\, d\epsilon\  \epsilon\ e^{-\epsilon/T}\over
\int_0^\infty\,d\epsilon\ e^{-\epsilon/T}}=T,
\label{1}
\eeq
which is {\em independent} of the frequency $w$. Integrating next over all
frequencies one encounters the ultraviolet catastrophe. It took a very drastic
and unheard off ``action" to fix this problem. Indeed, Planck suggested that
$\epsilon$ is not a continuous variable but a {\em discrete} one instead:
\beq
\epsilon=n(\hbar w), \quad n\in Z^+,
\label{2}
\eeq
where $\hbar\approx1\times10^{-34}({\rm J\cdot s})$ is a new dimensional
parameter. In such a case, Eq.~(\ref{1}) is replaced by
\beq
\vev{\epsilon}_T={\sum_{n=0}^\infty (n\hbar w) e^{-n\hbar w/T}\over
\sum_{n=0}^\infty e^{-n\hbar w/T}}={\hbar w\over e^{\hbar w/T}-1}\quad ,
\label{3}
\eeq
which depends on the frequency $w$. Integration over all frequencies yields
a finite result. The ultraviolet catastrophe has been averted. This bold
suggestion -- {\em quantize the energy} to resolve the black-body radiation
problem -- led to Quantum Physics, which is one of the two pillars of Modern
Physics in the 20th century.\\

\subsection{The Michelson-Morley Experiment}

One of the niceties of Newtonian mechanics is the incorporation of the
Galilean principle: all inertial frames are equivalent. The invariance of
Newtonian mechanics under the Galilean transformations:
$\vec x'=\vec x-\vec v t$, $t'=t+const$, with $\vec v=const$, imply the absence
of a preferred inertial reference frame and the existence of ``universal" time
or ``universal" simultaneity. To avoid problems with causality, action at a
distance or signals traveling with infinite velocity have to be incorporated.

On the other hand, electromagnetic theory as contained in Maxwell's equations:
\begin{eqnarray}
\vec\nabla\cdot\vec E&=& 4\pi\rho,\\
\vec\nabla\times\vec E&=&-{\partial \vec B\over\partial t},\\
\vec\nabla\cdot\vec B&=&0,\\
\vec\nabla\times\vec B&=&4\pi\vec j+{1\over c^2}{\partial\vec E\over\partial
t},
\label{4}
\end{eqnarray}
is not invariant under Galilean transformations! Let me take it a bit further,
since there is a multi-moral story here. Except for the last term
(${1\over c^2}{\partial\vec E\over\partial t}$) on the R.H.S. in Eq.~(\ref{4}),
everything else was known to Coulomb, Faraday, and
Ampere. It was Maxwell's brilliant contribution to suggest the existence of
this {\em extra tiny term} (with respect to all the other terms), undemanded
by experiments at that time. This {\em tiny term} changed everything:
\begin{enumerate}
\item \begin{enumerate}
\item The dimensional constant $c$, was found (by Maxwell) to be the velocity
of light ($\sim300,000$km/sec)!
\item The very existence of this dimensional constant ($c$) led eventually to
special relativity.
\end{enumerate}
\item \begin{enumerate}
\item A ``symmetric" treatment of $\vec E$ and $\vec B$, thus leading to
electromagnetism (E-M).
\item The prediction and subsequent discovery of electromagnetic waves.
\end{enumerate}
\item E-M theory, in quantum form, became the archetype of modern gauge
theories.
\end{enumerate}

It was precisely the assumption that the velocity of light $c$, {\em remains
constant}
in all inertial frames, that made Einstein abandon the Galilean transformations
and endorse the Lorentz transformations [$x'=(x-vt)/\sqrt{1-v^2/c^2}$,
$t'=(t-vx/c^2)/\sqrt{1-v^2/c^2}$, $y'=y$, $z'=z$] which eventually changed
profoundly our view of space and time. Indeed, if we had to insist on the
validity of Maxwell's equations, we had to abandon the Galilean relativity
principle, and assume the existence of a preferable reference frame, the
``ether-frame". The Michelson-Morley experiment showed that we ``ran out of
ether", thus vindicating the existence of some relativity principle which is
nothing else but special relativity: ``All laws of Physics are invariant under
Lorentz transformations". In this case, the Galilean invariants $\Delta
x=\Delta x'$, $\Delta t=\Delta t'$ are replaced by only one invariant:
$ds^2=c^2dt^2-d\vec x^2$, with far reaching consequences:
\begin{description}
\item (a) We should not talk any more about space and time, but about
space-time.
\item (b) There is no ``universal" time, thus the very idea of ``simultaneity"
becomes {\rm relative}! Past, present, and future depend on the ``eye" of the
moving observer, for space-like events. There is no ``action at a distance",
and signals cannot travel with infinite velocity; the velocity of light is the
maximal, limiting, allowable velocity. As the Minkowski line element ($ds^2$)
indicates, there is a horizon, beyond which there is no possible communication.
\end{description}

While points (a) and (b) changed profoundly our view of the Universe,
and (special) relativity is the second pillar of modern physics in the 20th
century, a few comments should be made:
\begin{description}
\item (i) In Einstein's Universe, events take place in a ``{\em pre-fixed}"
space-time: there is no arrow of time. After all, if time is another
dimension to be adjuncted to space, why should it have an {\em arrow}? Does
space have an {\em arrow}? Certainly not! Furthermore, notice that the Lorentz
transformations  are {\em effectively} two-dimensional, the transverse (to
$\vec v$) dimensions play no role! Then, recall the fact that in
two-dimensional physical theories there is no distinction between
$(x,t)\leftrightarrow(t,x)$ and if $x$ does not have an {\em arrow},
then why should $t$ have one? It does not have an {\em arrow}!
\item (ii) In accordance with the spirit of special relativity, all equations
of (classical and quantum) physics, are time-symmetric: they are invariant
under the transformation $t\leftrightarrow-t$.
\end{description}

This lack of an {\em arrow of time} in Einstein's Universe seems to be in some
conflict with common sense or ``conscious" time. After all, central to our
feelings of awareness is the sensation of progression of time: there is
a definite past and an uncertain future. Our perception of time seems to be in
severe discrepancy with the Einstein time of modern physics, where each
uniformly-moving observer has her own idea of simultaneity! There is no unique
concept of simultaneity and thus there is no definite past and undetermined
future, the whole space-time is fixed. How is it possible then to reconcile
the highly successful modern physical theory, based on Einstein's relativity,
and thus with no apparent arrow of time, with our gut-feeling of ``flowing"
time?

Somewhere something has to change!\\

\subsection{Line Atomic Spectra}

The idea of {\em Democritean} ``a-toms", gained progressively a lot of
support during the 19th century, despite the vicious attacks by many well known
figures of that epoch, \eg, Mach. Eventually it was realized that the classical
``solar system-like" picture of the atom \'a la Rutherford, led to severe
problems. Circulating electrons radiate energy according to E-M theory, and
thus eventually, unavoidably, collapse into the nuclear center. It took the
genius of Bohr, Heisenberg, Schr\"odinger, and Dirac to figure it all out and
led us to a completely new picture of the subatomic world, {\bf the quantum
world}, profoundly different from our classical view of reality. The notion
of {\em wave-particle duality} is the {\em central dogma} of Quantum Physics.
It is best expressed by Heisenberg's {\em uncertainty principle}
\beq
\Delta x\cdot\Delta p\ge \hbar,
\label{8p}
\eeq
which emphasizes our inability to ``measure" {\em simultaneously} the position
and momentum of a particle, as precisely as we want. A sharp deviation from our
classical view of the world! All the information about the system is embodied
in its complex wave-function $\Psi$, and its evolution is described by some
suitable wave-equation (Schr\"odinger, Klein-Gordon, Dirac, ...). An essential
point of quantum reality is the principle of {\em complex linear superposition}
or {\em quantum parallelism}, or {\em quantum coherence}, which is reflected
in the specific forms that the relevant wave equations take. Any number of
(quantum) states whatever, {\em irrespectively} of how different they are, can
coexist in any complex linear superposition. A ``measurement", or an
``observation" on the system, \ie, magnification of quantum effects at a
classical level, consists of picking up one of the actual alternatives,
$\Psi_i$, with classical probability $P=|\Psi_i|^2$, according to Born. This
arguably strange picture of the subatomic world, quantum reality, seems to
work perfectly well and constitutes together with special relativity, the
foundations of modern physics. It came a long way from Planck's bold assumption
of energy quantization and has led to the miracles of the quantum world.

Hopefully, by now I have convinced you that sometimes innocent looking ``small"
problems may carry the seed, if properly analyzed, of potentially big
discoveries, and may act as {\em time} bombs ticking at the foundations of well
established/accepted theories/views about the physical world.

\newpage
\section{Recent Past and Present}

Towards the end of the 20th century, physics seems to be in very good shape.
Quantum physics, in the form of Quantum Field Theory (QFT), has given us the
Standard Model (SM): an $SU(3)\times SU(2)\times U(1)$ gauge theory that
describes strong  ($SU(3)$) and electroweak interactions
($SU(2)\times U(1)$). All available experimental data are in remarkable
agreement with the SM. Despite its phenomenal success, many people are under
the impression that we have to go beyond the SM, at least for aesthetical and
philosophical reasons. Grand Unified Theories (GUTs) try to unify all above
interactions into one, Supersymmetry (SUSY) tries to establish a
fermion-boson symmetry, and Supergravity GUTs try to put the whole thing
together, including gravity. While these efforts are well justified and try to
answer some big questions, I will assume here that some ``effective
supergravity" GUT theory will bring peace to the minds of these daring souls.
My purpose is to concentrate on a few ``small" problems here and there,
which many physicists believe that the currently available ``technology" will
eventually resolve in a smooth fashion. On my list, the most pronounced
``small" problems are:
\begin{enumerate}
\item Black holes: the Hawking catastrophe.
\item Quantum physics and gravity: do not mix.
\item
\parbox{2.5in}{\hfill\\
Schr\"odinger's cat/ \\ Einstein-Podolski-Rosen (EPR) ``Paradoxes"}:\
\parbox{3in}{collapse of the wave-function \\
nonlocality}
\end{enumerate}

Let me discuss now each of these problems and their possible consequences.\\

\subsection{Black Holes}

According to classical general relativity, neutral Black Holes (BH) are stable
objects. Thanks to Hawking's remarkable insight \cite{4}, we know that black
holes radiate almost like a black body, when some form of quantum effects (in
the semiclassical approximation) are taken into account, thus becoming
unstable. A black hole of mass $M$ may be viewed as a thermodynamic system at a
temperature $T_{\rm BH}$ and entropy $S_{\rm BH}$, where \cite{4,5}
\beq
S_{\rm BH}\sim M^2\quad{\rm and}\quad T_{\rm BH}\sim 1/M
\label{6}
\eeq
in accordance with the standard thermodynamic relation $dM=TdS$.

As the black hole radiates, it loses energy and eventually it ``evaporates".
The lifetime of a black hole of mass $M$ is given by
\beq
\tau_{\rm BH}\propto M^3
\label{7}
\eeq
In Eqs.~(\ref{6},\ref{7}) I have used the so-called natural units: $k_B=1$,
$c=1$, $\hbar=1$, and $G_N$, the gravitational constant, which may be written
as $G_N=1/M^2_{Pl}$, has also been set to one.  The Planck mass is
$M_{Pl}\approx10^{19}\GeV$. While the above analysis sounds and looks like
standard thermodynamics, it contains the seeds of potentially profound and
dramatic changes, both in quantum physics and general relativity! The central
issue is {\em loss of information}. Imagine that our black hole can be
described, according to quantum mechanics, by a {\em pure state} $\Psi$. As it
{\em thermally radiates}, it evolves into a {\em mixed state}, which contains
much less information about the
black hole system, as compared to its initial state. A lot of information has
been lost, which is in agreement with the huge entropy that characterizes a
black hole (see Eq.~(\ref{6}))! After all, black body radiation is notoriously
universal and independent of the specific details of the radiating body, thus
clearly not the best form of information storage. Furthermore, this transition
from a {\em pure} to a {\em mixed} state is not allowed in quantum mechanics,
because it leads to a breakdown of our central {\em sacred quantum principle}:
{\em quantum complex linear superposition} or {\em quantum coherence}.
According to quantum mechanics, {\em purity is eternal}\ ! The black hole
evaporation, or {\em Hawking radiation} looks like an unprecedented attack at
the very basis of quantum physics: quantum coherence. That is why I call it the
{\em Hawking catastrophe}. Hawking himself was not slow to realize the
profound significance of his discovery and suggested that we may have to
abandon the principle of quantum coherence \cite{6}. A rather bold proposal! He
went further to suggest that we may have to abandon altogether the very notion
of the wave-function $\Psi$ to describe quantum states, and that we should
replace it with the notion of density matrix $\rho_{ij}(\sim\Psi_i\Psi^*_j)$.
This kind of dramatic change implies that we have to abandon the very idea of
the $S$-matrix: $\Psi_{\rm out}=S\Psi_{\rm in}$ and replace it by the $\S$,
the superscattering matrix \cite{6}:
\beq
\rho_{\rm out}=\S\rho_{\rm in}
\label{8}
\eeq
J.~Ellis, J.~Hagelin, M.~Srednicki, and myself \cite{7} took it a bit further:
if the
idea of the wave-function is to be abandoned, then the corresponding
wave-equation is to be abandoned and to be replaced by some form of equations
for the density matrix $\rho$. It was known, since the early thirties, that the
quantum wave equation could be cast in the form of Liouville-type equations,
similar to ones describing statistical systems, which involve directly the
density matrix: $\partial\rho/\partial t=i[\rho,H]$, where $H$ is the
Hamiltonian (evolution operator) describing our system. Our search for an
appropriate generalization of the above equation lead us to the following
{\bf M}aster {\bf E}quation (ME) \cite{7}:
\beq
{\partial\rho\over\partial t}=i[\rho,H]+\delta\H\rho.
\label{9}
\eeq
Here $\delta\H$ is some appropriate, system-dependent operator, which
basically renders the system ``open", and has to satisfy a lot of conditions
in order to be acceptable: conservation of probability, conservation of energy,
increasing or stationary entropy, etc. Very ``tall-orders" to realize in
practice, specifically if there is no consistent theory of quantum gravity.
Nonetheless, ``the cat is out of the bag" now, and the Hawking catastrophe
has to be addressed and resolved one way or another. It should be stressed
that our problem is at the fundamental level of the theory, and in a way,
independent of the existence of astrophysical black holes. At very
short distances, close to the Planck scale ($l_{Pl}\sim10^{-33}$~cm) the
space-time metric, like any other quantum field, fluctuates violently thus
destroying our ``fixed" space-time notions and leads to the so-called {\em
space-time foam} \cite{8}. There is no ``fixed" space-time at the Planck scale,
only {\em space-time foam} due to quantum fluctuations, a dynamical effect
beyond our control. Specifically, virtual black holes of Planck size certainly
contribute to the space-time foam. As a particle moves through space-time foam,
it encounters, in principle, these Planck-size black holes, which suck
information and ``evaporate", thus ``opening" the system (a particle in our
example). This is the ``microscopic" origin of the extra term $\delta\H\rho$
in Eq.~(\ref{9}), which sums up the effects of space-time foam on the system
and it turns it, {\em unavoidably}, into an ``open" system.

While intuitively plausible, our analysis above has been criticized on the
following grounds: Hawking's radiation is based on some semiclassical
approximation, in lack of a complete theory of quantum gravity, and thus
should not be taken extremely seriously, specifically if it runs against
big principles such as quantum coherence. It may all be some artifact of our
unjustifiable approximations. This criticism, sound for the time being, brings
me to another ``small" problem.\\

\subsection{Quantum Physics and Gravity}

All efforts to extend our highly successful unified theories of gauge
interactions (strong and electroweak) to include gravity into a consistent
quantum dynamical framework, have badly failed, including supergravity
theories. The problem being our inability to tame all kinds of infinities
that emerge at each order in perturbation theory in sharp contrast with
renormalizable gauge theories. While the standard gauge interactions are
characterized by dimensionless coupling constants (\eg, electric charge), the
gravitational constant $G_N$ is dimensional, thus preventing the usual
renormalization program to apply. The theory has to be directly {\em finite},
no infinities are allowed because, if present, they are incurable. It should be
stressed that this problem seems to bring us against a brick wall, since we
have used all kinds of symmetries that we can envisage: space-time symmetries,
internal symmetries (local or global), fermion-boson symmetry or supersymmetry
(local or global), and still no acceptable quantum theory of gravity in sight.

This is a rather serious and threatening impasse.\\

\subsection{Schr\"odinger's Cat/EPR ``Paradoxes"}

While quantum physics seems to describe extremely accurately
the microworld, its ``blind" extension to the macroworld may lead to rather
remarkable ``paradoxes", or even nonsensical answers. Quantum parallelism or
quantum coherence, {\em if} assumed valid all the way to the macrocosmos,
implies the coexistence of vastly different macroscopic states, in sharp
contrast with common sense, everyday experience. A celebrated example of
applying quantum physics all the way is Schr\"odinger's cat \cite{9}. In an
``isolated" box, a cat coexists with a vial of poison and some lump of
radioactive material, such that after some time $\tau$ there is 50\%
probability that at least one nucleus has decayed, and by triggering a series
of well-defined processes results in the breaking of the vial of poison and
thus the death of the cat. According to standard Quantum Mechanics (QM), the
system (cat + nucleus) is described by:
\beq
\Psi_{\rm system}(t)={1\over\sqrt{2}}\Psi^{\rm alive}_{\rm cat}\Psi^{\rm
stable}_{\rm nucl}+{1\over\sqrt{2}}\Psi^{\rm dead}_{\rm cat}\Psi^{\rm
decayed}_{\rm nucl},
\label{10}
\eeq
for $t\ge\tau$.

Before making any ``measurement" (\eg, opening the box), the system is
described by Eq.~(\ref{10}), obeying standard quantum mechanics equations.
Thus we have a certain moment where the cat is 50\% dead and 50\% alive, etc.
Even if cats have ``nine lives", this is a kind of result difficult to
swallow! Here we have an archetypal example of a microprocess (nuclear decay)
perfectly described by QM, leading to a macroscopic {\em logical catastrophe},
if we insist that our combined macro-micro system is {\em QM-describable}!
Of course, the crucial point here is our assumption that a macrosystem
(\eg, our cat) obeys the laws of quantum physics. But, on the other hand, the
whole point of {\em reductionism}, the {\em tacit} dogma of modern physics, is
that everything can be understood in terms of few elementary constituents and
their interactions. So, if our cat is made of these elementary constituents,
who is there to decide when and how the transition from the micro to the macro
system occurs, in order to avoid situations where the cat is 50\% alive and
50\% dead? Ideally, what we are after is some {\em dynamical mechanism} that
destroys quantum coherence for macroscopic systems by {\em spontaneous collapse
of the wave function}, with a realistic time scale compatible with our notions
of the macrocosmos. Here we have touched upon a rather big issue of quantum
physics, that of the {\em collapse of the wavefunction} which presumably occurs
when a ``measurement" takes place. By its very nature, a ``measurement"
involves a macrosystem in ``contact" with the ``measured" microsystem, that
triggers the microsystem to {\em jump} into one of its possible quantum states
$\Psi_i$ with classical probability $|\Psi_i|^2$. As is well known, this
``quantum jump" process has made many people very skeptical, including
Schr\"odinger, of whether quantum physics in its standard-textbook form is the
whole story. The dynamics of the {\em collapse of the wavefunction} do not seem
to be completely and satisfactorily understood.

Another example exposing some of the possible inadequacies of standard quantum
physics is the famous Einstein-Podolsky-Rosen (EPR) ``paradox" \cite{9}.
Consider a spin-0 system (\eg, a $\pi^0$) decaying at rest into two photons,
moving in opposite directions (conservation of linear momentum). Clearly, the
total angular momentum of the system of the two photons has to be always zero.
The wave function of the system of the two photons can be represented by:
\beq
\Psi_{\rm system}\propto (\gamma_1)_+(\gamma_2)_-+(\gamma_1)_-(\gamma_2)_+
\label{11}
\eeq
where $+(-)$ are the helicities of the photon with respect to some
(unspecified) direction. Before making any ``measurement" on the system, \eg,
measuring the helicity of the photon $\gamma_1$, the question ``what is the
helicity of any one of the photons?" is an empty one. The two photons are
entangled in such a way that we can only describe them as lumped together
in $\Psi_{\rm system}$. Imagine now that some time after the parent decay, we
``measure" the helicity of photon $\gamma_1$ and we find it to be $(-)$, then
we know, according to Eq.~(\ref{11}), with {\em certainty} that if we
``measure" next the helicity of $\gamma_2$, it is going to be $(+)$. What has
happened is that our first ``measurement" has demolished/collapsed the
$\Psi_{\rm system}$ down to the second term, $(\gamma_1)_-(\gamma_2)_+$:
\beq
(\gamma_1)_+(\gamma_2)_-+(\gamma_1)_-(\gamma_2)_+
\begin{array}{c}
{\rm ``measurement"}\\
\longrightarrow\\
{\rm ``jump"}\\
\end{array}
 (\gamma_1)_-(\gamma_2)_+
\label{12}
\eeq
While innocent looking, the above standard lore of quantum mechanics leads to
some dramatic consequences. After the parent decay, the two photons go their
ways and clearly they are well separated (they are not both at the same
space point, for all times). How is it possible then, that the photon
$\gamma_2$ knows ``instantenously" the direction of its helicity, according to
Eq.~(\ref{12}), \ie, at the moment that the $\gamma_1$ helicity is measured?
Apparently we have to attribute some {\em nonlocality to the quantum jump}!

It should be {\em emphatically stressed} that the above analysis should not be
taken to imply that we can use EPR-type experiments to send signals faster than
light! The {\em nonlocal influences} that arise in EPR-type experiments are
not such that they can be used to send messages. In our particular example
above, it is only the direction of the alternative polarizations which arrives
faster than light (``instantenously"), while which of the two possible
directions ($+$ or $-$) has been picked up, arrives through ``normal channels"
of communicating the result of the first ($\gamma_1$) polarization measurement.
So, the nonlocal nature of the quantum jumps does not lead to any violation of
causality! Actually, there is a whole new activity on the {\em basics} of
quantum mechanics, due mainly to the ingenuity of John Bell, who through his
inequalities established that no {\em local ``realistic"} (\eg, ``hidden
variable", or ``classical type") description can give the correct quantum
probabilities \cite{9}. Amazingly enough, a detailed series of experiments in
the mid 1980's, performed by A. Aspect and collaborators \cite{10}, have {\em
vindicated} standard quantum mechanics, \ie, {\em the nonlocal nature of
quantum jumps}, ruling out {\em local, realistic models}.

As far as we do {\em understand} and {\em accept the nonlocal nature of
quantum physics}, there is no EPR-``paradox". Alas, we have not finished yet.
As has been emphasized repeatedly by Penrose \cite{9}, all the above analysis
of the EPR-``paradox" and its satisfactory resolution in the framework of
quantum mechanics is {\em non-relativistic in nature}. As I mentioned above,
the two photons are well space-separated, thus the two measurements  (measuring
the helicity of $\gamma_1$ and then measuring the helicity of $\gamma_2$)
correspond to space-like separated events. As such, as I emphasized in section
1, in accordance with special relativity, which measurement occurs first
depends on the ``eye" of the moving observer! Imagine that $\gamma_1$ moves
to the left and thus $\gamma_2$ to the right. For an observer moving  rapidly
enough to the left, she will tabulate the events as follows: (1) measurement
of the $\gamma_1$ polarization and quantum jump, (2) measurement of the
$\gamma_2$ polarization. While for and observer moving rapidly enough to
the right, it looks like: (1) measurement of the $\gamma_2$ polarization and
quantum jump, (2) measurement of the $\gamma_1$ polarization. What is going on?
There is some real conflict here: we are getting {\em two different pictures
of physical reality}! Different measurements cause the nonlocal jump. According
to Penrose \cite{9}, there is an essential conflict between our space-time
picture of physical reality, including the nonlocal nature of quantum physics,
and the spirit of special relativity. Let me appropriately call it the {\em
Penrose paradox} \cite{9}.

I am not aware of any better example showing so graphically the apparent
inadequacies at the very foundations of QM and special relativity, in
providing a complete and satisfactory picture of physical reality. The very
issue of the apparent lack of ``universal" simultaneity (or ``universal" time)
in special relativity, combined with the apparent nonlocal nature of the
quantum jumps (``collapse of the wavefunction") drive our sense of physical
reality against the wall! While this state of affairs is rather upsetting and
alarming, it indicates the profound intimacy between the very nature of time
and the collapse of the wavefunction. A possible resolution of the Penrose
paradox \cite{9} should definitely try to shed light on this close intimacy.

Actually, we shouldn't be that surprised by decoding the ``message" contained
in the Penrose paradox \cite{9}. After all, the very notion of the collapse of
the wavefunction is, for at least a few people, a {\em time asymmetric
process}. Consider for example, after Penrose \cite{9}, a lamp emitting
photons, which go through
a half-silvered mirror, tilted at $45^\circ$ with respect to a horizontal line
that connects the lamp and a photocell. Clearly, the probability that the
photocell detects a photon, {\em given} that the source emits one, is exactly
$(1/\sqrt{2})^2=1/2$, so that 50\% of photons are absorbed by the wall.
A straightforward application of the standard QM rules, employing {\em
time-reversal symmetry}, will give the same answer (1/2) for the probability
that the lamp emitted a photon, {\em given} that the photocell detected a
photon! Sheer nonsense! The {\em correct} experimental answer is of course 1.
There are no off-wall photons registered by the photocell! This example shows
dramatically that the ``collapse of the wavefunction" is not as innocent as
it looks, \ie, just the amplitude squaring procedure, but something more sneaks
into the whole process. It is hard to avoid thinking of some kind of {\em
thermodynamic irreversibility}, implying some {\em entropy increase}, thus
defining clearly {\em an arrow of time}, and thus turning the ``collapse
of the wavefunction" into {\em a non-relativistic process}!

Once more,  somewhere, something has to change!

\section{Interlude (I) -- Free Time}

It will be instructive to try to isolate the {\em possible common origin} of
the three ``small" problems discussed in the previous section. Let us start
with the Hawking catastrophe, endemic to black holes. The central issue here
is the apparent loss of information, as reflected in the huge entropy (see
Eq.~(\ref{6})) characterizing black holes. The problem being that black holes
have ``{\em no hair}": there are not many useful, suitable quantum numbers
that may label the black hole. Mass ($M$), angular momentum ($\vec L$), and
electric charge ($Q$) are about the only ``charges" that are measurable
at {\em long distances}, beyond the black holes's horizon, due to long-range
forces (gravitational, electromagnetic). These long-range ``charges" are
scarcely enough to describe completely the state of a black hole, containing
say, at least $3M_\odot ({\rm grams})N_{\rm Avogadro}$ basic constituents.
In other words, there is a large number of different configurations, all
characterized by the same $M$, $\vec L$, and $Q$! That is why we are getting
such a huge entropy $\propto M^2$. A possible cure for this situation is to
invent more kinds of suitable ``charges". But we know that there are not that
many new long-range interactions remaining to be discovered! Furthermore, the
above discussion shows that we {\em practically} need an {\em infinite number}
of new, suitable ``charges". Another apparent impasse? Not necessarily! Thanks
to quantum mechanics it is possible to measure a specific kind of ``charge",
let me call it the Aharonov-Bohm (AB) ``charge", at long distances, even if
these ``charges" are not necessarily due to long-range forces. We may make
use of the celebrated AB effect \cite{11}, where \eg, by measuring the phase
shift that an electron (wavefunction) suffers when circulating around (but at a
long distance outside) a solenoid, we may deduce the flux (solenoid's AB
``charge") and thus the magnetic field {\em inside} the solenoid, which as is
well known, vanishes outside the solenoid. For example, in a double-slit
experiment, the existence of an ``ideal" solenoid between the two slits and the
screen will shift the total wave pattern on the screen by an amount
\beq
(\Delta\theta)_{\rm AB}={Q\Phi\over\hbar},
\label{13}
\eeq
with $Q$ denoting the electric charge of the projectile and $\Phi$ the flux
through the solenoid, which may be considered as solenoid's AB ``charge".
If this type of AB ``charges" were available for a black hole, then by
allowing suitable ``projectiles" to circulate around (but far!) from a
black hole, we could in principle measure all of its ``charges". If a black
hole carries AB ``charges", because of its quantum nature, it is called {\em
quantum hair} \cite{12}. While the need for a huge number of new long-range
interactions has evaporated, the source of a huge number of different kinds of
quantum hair remains enigmatic. A new impasse in the (BH) horizon! In this case
we have come close to the brink. All our symmetries, space-time, internal,
supersymmetry, have been exhausted and still we have no answer! Take, as we
have {\em tacitly assumed, a point-like} particle: you can accelerate it ($m$),
kick it around ($\vec p,\vec L$), give it some internal ``charges" (electric,
color, ...), ``change" its spin (supersymmetric charge), make it ``live" in
a D-dimensional Universe with the D-4 extra dimensions curled up \'a la
Kaluza-Klein, and still, we have a long way to go before
its I.D. contains a practically infinite kind of lines, each for every
(spontaneously broken) symmetry (AB ``charge"). After all, in every
point-like quantum field theory we have a number(usually small) of elementary
fields participating in a finite number (usually small) of interactions,
thus falling extremely short from the practically needed infinite amount of AB
``charges", for a complete description of a black hole. Of course, the
situation is dramatically different if we abandon the idea of point-like
particles as fundamental blocks and use instead {\em extended objects}. In
such a case, the particle spectrum is going to be {\em dynamically} infinite
(corresponding to all kinds of vibrational or other modes), thus providing the
seeds for a practically infinite number of AB ``charges", corresponding  to an
infinite number of spontaneously broken generalized ``gauge" symmetries due to
the availability of an infinite number of generalized ``gauge" bosons.

Actually, the above analysis of the cause of information loss in black holes,
\ie, the {\em no-hair theorem}, and its possible resolution by abandoning the
{\em point-like nature} of elementary constituents and moving to {\em
elementary} but {\em extended} fundamental blocks, applies equally well to
our second ``small" problem: no consistent theory of quantum gravity. Indeed,
it is well known that most of the infinities that plague quantum field theories
are due to the assumed point-like nature of the elementary fields-particles.
A mere glimpse at Newton's gravitational law, or Coulomb's law, at very short
distances, exhibits an ``apparent" infinity as $r\to0$! As described in the
previous section, in the case of Coulomb's law, as contained in quantum
electrodynamics, or more generally in gauge theories, we do have an algorithm,
called {\em renormalization} to take care of all types of infinities, but
apparently not so for gravity. It seems that our tacit assumption of
{\em point like elementary constituents} is, again,  the stumbling block for a
consistent quantum theory of gravity. The move to {\em elementary} but {\em
extended} fundamental blocks, looks once more unavoidable!

Concerning the Schr\"odinger's cat and the EPR/Penrose ``paradoxes" \cite{9},
\ie, {\em the collapse of the wavefunction}, involving {\em nonlocal quantum
jumps} of {\em non-relativistic nature}, a similar analysis applies as to
the other two ``small" problems above. Indeed, our assumption that the
principles of quantum physics, mainly quantum coherence, applies all the way
from the Planck scale to the cat-scale, is questionable. It should be that
some {\em dynamics} intervene in the way that make it highly improbable for the
cat to be described by a coherent wavefunction. But what can it be? Well,
maybe we can, very naively, put our first ``small" problem, the BH-Hawking
catastrophe, to work to our advantage! Everybody knows that at Planck
distances, quantum fluctuations of the metric destroy the {\em finesse}
of our space-time and create space-time foam. Virtual, Planck scale BHs appear
and disappear continuously, sucking information from our ``system" (see
Eq.~(\ref{9})). Clearly, these (quantum) gravitational effects are proportional
($(G_NE^2)^n$, n=1,2,...) to the mass (or maybe energy) of our ``system". Thus,
it is not inconceivable that ``elementary systems" (\eg, electrons) remain
virtually unscathed by the Planck BH-Hawking catastrophe, while ``complex
systems" (\eg, our cat) suffers a {\em spontaneous} collapse of their
wavefunction. So an {\em effective Planck BH-Hawking catastrophe}, may be
after all welcomed from the collapse of the wavefunction point of view. The
word {\em effective} is of dramatic importance here: we discussed above
that {\em point-like} QFT type of BH are unacceptable, too many profound
problems, and we speculated that {\em extended objects} QFT type of BH may be
the solution. The extra demand here is that this new type of BH dynamics,
based on {\em extended objects QFT}, should allow in some appropriate limit
for some controllable leakage of information, thus mimicking {\em effectively}
the Planck BH-Hawking catastrophe and triggering the collapse of the
wavefunction! Incidentally, this {\em effective Planck BH-Hawking catastrophe}
triggering {\em spontaneous} collapse of the wavefunction, meets successfully
the {\em demand}, discussed in the previous section, that some kind of
thermodynamic irreversibility, accompanied by entropy increase, is involved
somewhere in the ``collapse of the wavefunction" process.

The need for {\em extended}, {\em fundamental objects}, as opposed to {\em
point-like constituents}, for the collapse of the wavefunction is enforced by
the EPR-Penrose ``paradox", involving nonlocal quantum jumps presumably of
non-relativistic nature. The very straight connection between causality and the
vanishing of any space-like correlations in point-like QFT, becomes much looser
in the case of extended objects. After all, the infinitely sharp
delta-functions, involved in the definition of space-like events, are smeared
out over the extension of our fundamental objects, thus making it, at least
intuitively plausible, for the quantum mechanical ``instantaneous" influences,
as observed in the EPR-type experiments, \eg, by A. Aspect \etal\ \cite{10}.
Furthermore, the very existence of {\em extended, fundamental objects}, entails
the existence of some fundamental length, thus a {\em new dimensional
``constant"} of nature. It is far from clear that our notions of special or
general relativity based on point-like constituents (practically equivalent to
the fact that $c$, the velocity of light, is the {\em only dimensional
parameter} at the classical level), will hold true in the case of {\em
extended, fundamental objects} (practically introducing a new fundamental
dimensional ``constant": their size). Past experience shows (see Section 1)
that moving from classical mechanics (no dimensional constants) to Maxwell's
equations (E-M theory) (one dimensional constant: $c$) and asking for some
common sense relativistic principle to hold true, shook dramatically the
foundations of classical physics, specifically the very notion of {\em time}!
So, it does not look inconceivable that we may be in for another ``shake up"
of our assumed solid foundations of modern physics, and specifically once more
involving the notion of time. It is amusing to note that the very existence of
a fundamental length, if interpreted as the modern version of the ``ideal rigid
body" of classical mechanics, defies the very notion of ``relative
simultaneity"! Its ends would always move simultaneously as observed from any
frame, and it could therefore be used to establish ``universal time"! It may be
that special (or general) relativity is some {\em appropriate limit} of a more
fundamental theory, and thus {\em approximately true} even if, in lots of
circumstances, this approximation is {\em extremely} accurate! Thus, the hope
exists that the dynamics of extended fundamental objects may lead to some new
notion of {\em time}, encompassing the possibility of ``universal
simultaneity", and allowing for space-like correlations, thus evading the
EPR-Penrose paradox.

Before moving to the discussion of the simplest fundamental extended objects,
\ie, superstrings, it would pay to put forward what exactly are we expecting
from a {\em complete theory} of {\em quantum physics}. Until now, our approach
to the quantum world involves two components: the one component dubbed by
Penrose \cite{9} the {\bf U}-part, involves the {\bf U}nitary evolution of the
system, in a {\em deterministic}, {\em continuous}, {\em time-symmetric}
fashion, as described for example by the Schr\"odinger equation,
\beq
{\bf U}:\qquad i\hbar{\partial\Psi\over\partial t}= H\Psi,
\label{14}
\eeq
with $H$ the Hamiltonian operator of our system described by the
wavefunction $\Psi$. Clearly such an evolution respects quantum coherence,
as reflected by the quantum complex superposition principle implicit in
Eq.~(\ref{14}). The second component, dubbed by Penrose \cite{9} the {\bf
R}-part, involves the {\bf R}eduction of the state-vector or collapse of the
wavefunction, that enforces coexisting alternatives to resolve themselves
into {\em actual} alternatives, one {\em or} the other,
\beq
{\bf R}:\qquad \Psi=\sum_i c_i\Psi_i\longrightarrow \sum_i|c_i|^2|\Psi_i|^2,
\label{15}
\eeq
where $|c_i|^2$ are classical probabilities describing {\em actual}
alternatives. It is the {\bf R}-part of quantum physics that introduces
``uncertainties" and ``probabilities", thus involving {\em discontinuous,
time-asymmetric quantum jumps} and leading to gross violations of quantum
coherence. It is fair to say that almost universally, when physicists talk
about quantum physics, they tacitly identify it with its {\bf U}-part only! It
is the {\bf U}-part that has absorbed all our attention for about 70 years now,
and in its more advanced form, relativistic quantum field theory, has become
an icon of modern physics, with spectacular success, \eg, the standard
model $SU(3)\times SU(2)\times U(1)$. On the other hand, the {\bf R}-part has
been vastly and conveniently forgotten, tacitly assumed to be some mere
technicality that gives us the right rules of ``measurement" or ``observation":
different aspects of a quantum system are simultaneously magnified at the
classical level, and between which the system must choose. I strongly believe
that this attitude has brought us finally to a dead end, \eg, the {\em Penrose}
paradox, and we need to reconsider our strategy. Actually, I believe that,
maybe against the ``mainstream", there is no way to deduce the {\bf R}-part
from the {\bf U}-part, the {\bf R}-part
being a completely different procedure from the {\bf U}-part, and effectively
providing the {\em other ``half"} of the interpretation of quantum mechanics.
It is the ({\bf U}+{\bf R}) parts {\em together} that are needed for the
spectacular agreement of QM with the observed facts. So, we are after some {\bf
N}{\em ew dynamics}, {\bf N}, that provides a unified and comprehensible
picture of the whole ({\bf U}+{\bf R}) process, by giving satisfactory answers
to basic questions like: what constitutes ``making a measurement"? Why and when
are $|\Psi_i|^2$ interpretable as probabilities? How does the classical world
emerge from the quantum world? Let me schematically represent the above
approach by
\beq
{\bf U}\oplus {\bf R}\subseteq {\bf N}
\label{16}
\eeq
It should be stressed that the {\bf N}ew dynamics involved in the {\bf
N}-equation (\ref{16}), because they have to approach at appropriate limits the
{\bf U}-equation (\ref{14}) and the {\bf R}-equation (\ref{15}), \ie, almost
anti-diametrical points of view, cannot be some smooth generalization of
some wave dynamics. Apparently, the {\bf N}-equation (\ref{16}) has to contain
seeds of {\em non-relativistic invariance} and {\em time asymmetry}, but in
such a way that when the {\bf R}-part or emergence of classicality is {\em
neglected},
an approximately relativistic, time-symmetric (quantum field) theory emerges.
{\em Neglect} in my thinking means that either we disregard completely the
notion of space-time foam, \ie, back to the ``fixed" space-time, {\em
topologically smooth} picture, {\em or}, in a more realistic way, we include
the effects of space-time foam and prove {\em dynamically} that at large
distances (compared to the Planck scale $l_{Pl}\sim10^{-33}$cm) and for
micro-objects (electrons, ...) an approximately relativistic time-symmetric
(quantum field) theory springs out.

Let us move next to the discussion of the dynamics of superstrings, the
simplest {\em extended}, {\em fundamental} objects, that are relevant to
our program here, to develop a possibly complete picture of the quantum
world. It should be stressed up front that superstrings are not only the
{\em simplest/extended, fundamental objects}, but also the only ones whose
dynamics have not been marred with grave problems.

\section{Present}

Superstring Theory (ST) \cite{13}, the theory of {\em one-dimensional extended
fundamental objects}, is the {\em only} theory known to provide a consistent
framework for Quantum Gravity. The particle spectrum of {\em closed strings}
contains {\em always} a massless spin-2 field that has all the right properties
to be identified with the {\em graviton}, the carrier of the gravitational
force. This remarkable property, combined with the natural existence in the
closed string spectrum of massless spin-1 fields, identifiable as the
mediators of the other forces (strong and electroweak), justify the singular
excitement that superstring theory has created, as a prominent candidate for
the {\bf T}heory {\bf O}f {\bf E}verything ({\bf TOE}). The fundamental length
of the string $l_{\rm string}$ is, as naturally expected in any theory of
Quantum Gravity, determined by the gravitational constant
\begin{equation}
l_{\rm string}\approx {\cal O}(\sqrt{G_N})\approx{\cal O}\left({1\over M_{\rm
Pl}}\right)
\approx{\cal O}(l_{\rm Planck})\approx {\cal O}(10^{-33}\cm).
\label{20}
\end{equation}

Fundamental strings have indeed minuscule extensions, and for many practical
purposes they can be safely assumed to be almost point-like. It is remarkable
and non-trivial that the low-energy or infrared limit of string theory provides
a {\em realistic}, {\em effective} supergravity, point-like quantum field
theory, which is able to describe the low-energy (with respect to the Planck
scale) world. Let me now describe some of the dynamics of string theory that I
will need for my main purpose here, \ie, to develop a new stringy-based quantum
physics framework, which hopefully will be able to satisfy all the ``{\em
desiderata}" set up in section 3.

\subsection{General string framework}
As a closed string moves in spacetime it sweeps out a two-dimensional surface,
{\em world-sheet}, which is described by two variables, $\sigma\in[0,\pi]$ and
$\tau$ (time) which runs on the real axis. It is very convenient to replace
$(\sigma,\tau)$
by a complex variable $z=e^{i\sigma+\tau}$ which maps an initial state at
$\tau_i=-\infty$
to $z=0$ and turns ``time-ordered" products into ``radially-ordered" products.
This
``complexification" is of extreme technical importance, since it leads to
the well-known
theory of Riemann surfaces and complex analysis. At the classical level, we
have to study
``physics" on the Riemann sphere, while quantum corrections (loops) correspond
to more
complicated Riemann surfaces, torus, etc. of genus ($\equiv$handles on the
sphere)
$g>0$. The use of just the complex plane as the world-sheet makes a lot of the
fundamental properties of string dynamics, like {\em conformal invariance},
easier to
grasp, thanks to the simple fact that any analytic function of $z$ corresponds
to a
conformal mapping on the $z$-plane! Clearly, the dynamics of one-dimensional
extended
fundamental objects is bound to be pretty constrained and complicated. After
all, it is
almost like treating an {\em infinite set of point-like particles} all at once!
Writing down
the complete {\em action} for a string moving in $D$-spacetime dimensions is a
horrendous
thing. Usually, one writes down an {\em effective action} in $D$-dimensions,
that contains
a lot of information relevant to low-energy physics and, depending on the level
of
sophistication, tries to encompass as many novel stringy properties as
possible.
While this approach is pretty good for ``normal" type of particle physics, it
is not good for
us, because we are after {\em exactly} the properties that differentiate
between a string
theory from point-like QFT. Nevertheless, there is an equivalent way to discuss
string
physics. Concentrate first on the ``physics" of the string world-sheet, and
translate afterwards
to the spacetime language. In other words, we have to work with two-dimensional
$(\sigma,
\tau)$ QFT, which is simpler than $D$-dimensional QFT! One considers then a set
of
world-sheet
bosonic fields $X^\mu(\sigma,\tau)$, $\mu=0,1,\ldots,D-1$, corresponding to
the
$D$ spacetime coordinates of our space-time ($X^0$), that in the string jargon
is called
{\em target space}. The {\em action}, consistent with two-dimensional
reparametrization
invariance and renormalizability, is given by
\begin{equation}
S_{\rm 2-d}=-{T\over2}\int d^2\sigma\sqrt{h} h^{\alpha\beta} g_{\mu\nu}(X^\mu)
\partial_\alpha X^\mu\partial_\beta X^\nu
\label{21}
\end{equation}
where $T$ is the string tension, sometimes written as
$T\equiv{1\over2\pi\alpha'}=
{\cal O}(1/l^2_{\rm string})$, $h^{\alpha\beta}$ is the 2-d
($\alpha,\beta=1,2$) world-sheet
metric tensor, and $g_{\mu\nu}=g_{\mu\nu}(X)$ is the {\em target space} metric
(symmetric
and traceless) tensor, which upon quantization provides the {\em graviton}.
Notice that
the action (\ref{21}) has several worth discussing {\em symmetries}.
\begin{description}
\item (A) {\em Local}: beyond the standard 2-d reparametrization invariance,
there is
{\em conformal invariance}
\begin{equation}
h_{\alpha\beta}\to e^{\phi(\sigma,\tau)}h_{\alpha\beta}
\label{22}
\end{equation}
with the {\em Liouville field}, $\phi(\sigma,\tau)$, decoupled at least at the
classical level!
It should be {\em emphatically} stressed that the {\em conformal invariance}
of the action
(\ref{21}) is of fundamental importance and it should remain valid {\em not
only} at the
classical level (where it is automatic) but to all orders in $T$ or $\alpha'$.
\item (B) {\em Global}: reflect the symmetries of the {\em background},
$g_{\mu\nu}(X)$,
in which the string is propagating. For example, if we take
$g_{\mu\nu}(X)=\eta_{\mu\nu}$,
\ie, the Minkowski metric, then {\em Lorentz}, or more generally, {\em Poincare
invariance}
emerges, as a {\em global} ($\sigma,\tau$-independent) 2-d symmetry,
corresponding to
a {\em local} ($X^\mu$-dependent) target space symmetry!
\end{description}

The action (\ref{21}) should be seen as  an action describing the dynamics of a
bunch of 2-d
(world-sheet) fields $X^\mu(z)$, corresponding to our real spacetime
coordinates, while
the ``coupling constants" $g_{\mu\nu}(X)$ of this {\em non-linear
$\sigma$-model} should
be seen as corresponding to physical (spacetime) fields, \eg, the graviton in
this particular
case. The action (\ref{21}) is a {\em very small part} of the complete 2-d
action, able to
describe, say, the Standard Model. Every physical field (particle) $g(X)$ will
appear
multiplying a {\em vertex operator} $V_g$, \eg, $\partial_\alpha
X^\mu\partial_\beta X^\nu$
for the graviton $g_{\mu\nu}(X)$, of suitable form such that it successfully
and uniquely
represents the particle on the world-sheet physics world. To complete our story
we have
to know how to derive the equations of motion (E.O.M.) for our physical
(\ie, target
space) fields and how to calculate scattering amplitudes or, through the usual
LSZ-trick,
correlation functions in target space.

\subsubsection{String equations of motion (E.O.M.)}
A characteristic property (for some, the defining property) of {\em conformal
invariance}
is the vanishing of the trace of the energy-momentum tensor describing our
theory.
For example in the case of the action (\ref{21}), with
$g_{\mu\nu}=\eta_{\mu\nu}$, the
corresponding energy-momentum tensor is
\begin{equation}
T_{\alpha\beta}=\partial_\alpha X^\mu\partial_\beta
X_\mu-{1\over2}h_{\alpha\beta}
h^{\alpha'\beta'}\partial_{\alpha'}X^\mu\partial_{\beta'}X_\mu,
\label{23}
\end{equation}
which is automatically traceless, $h^{\alpha\beta}T_{\alpha\beta}=0$, as a
consequence
of conformal invariance (Eq.~(\ref{22})). The above result is valid at the
classical level.
When quantum corrections are taken into account, there are going to be extra
terms
in (\ref{23}) and extra care has to be taken such that the tracelessness
property (equivalent
to conformal invariance) remains still valid. In principle, quantum corrections
introduce
infinities that have to be removed through the renormalization programme, which
one
way or the other introduces a {\em scale} into the problem and thus breaks
conformal
(or scale) invariance. What we are after is a {\em finite}, not just {\em
renormalizable},
2-d {\bf C}{\em onformal} {\bf F}{\em ield} {\bf T}{\em heory} (CFT), which
implies that all the
$\beta(g)$-functions
corresponding to the ``couplings" $g(X)$ (our physical target space fields)
better be
zero to all orders in perturbation theory. The vanishing of the
$\beta(g)$-functions
implies correspondingly the vanishing of a set of equations involving the
$g(X)$ ``couplings",
which better be our E.O.M. (we cannot have two different sets of equations
satisfied by the same fields) \cite{14}
\begin{equation}
\beta_{g_i}(g)\vert_{\rm 2-d\ world-sheet}=\left.
{\delta S_{\rm eff}\over\delta g_i}
\right\vert_{\rm D-Target-space}=0
\label{24}
\end{equation}
Equation (\ref{24}) is a {\em highly remarkable} equation: it relates a {\em
basic} property
of world-sheet physics ({\em 2-d conformal invariance}) to the variation of
the
{\em effective} target-space action w.r.t. the physical fields, \ie, the target
space E.O.M.!
It should be stressed, {\em at the outset}, that away from the conformal {\em
fixed point}
(\ie, where all $\beta_{g_i}(g)=0$) a more general form of dynamics
applies and, as we will discuss later, extra care is needed for its
interpretation in terms of
target-space physics.

\subsubsection{String target-space scattering amplitudes}
Using the language of {\em vertex operators} (discussed above) to describe
physical
fields on the world-sheet, it is not difficult to motivate the form that the
target-space scattering amplitude for $N$-particles takes \cite{13}
\begin{eqnarray}
A(g_1,p_1;\ldots;g_N,p_N)&=&\kappa^{N-2}\int {\cal D}X(\sigma,\tau)
{\cal D}h_{\alpha\beta}(\sigma,\tau)\, \exp(-S_{\rm 2-d})\cdot\nonumber\\
&&\cdot \prod_{i=1}^N V_{g_i}(p_i)
=\VEV{\prod_{i=1}^N V_{g_i}(p_i)},
\label{25}
\end{eqnarray}
with $V_{g_i}(p_i)\equiv\int d^2\sigma \sqrt{h}
V_{g_i}(\sigma,\tau)e^{ip_i\cdot X}$ the
emission or absorption operator of a string state of type $g_i$ and momentum
$p_i$,
and $\kappa$ is a coupling constant. This {\em Golden Rule} of string theory is
another
highly remarkable relation, carrying us again from complicated $D$-spacetime
scattering
amplitudes to simple 2-d world-sheet correlation functions $\VEV{\prod_{i=1}^N
V_{g_i}(p_i)}$. Once more, it should be stressed that away from the conformal
fixed
point extra care should be taken in using Eq.~(\ref{25}) because, as we will
see later,
more general dynamics set in which may obscure its straightforward physical
interpretation.

\subsection{Critical versus non-critical string theory}
String theory as a consistent theory of gravity should be able to take care of
two
fundamental problems:
\begin{description}
\item (A) Quantum Gravitational (QG) corrections to scattering processes, {\em
in
fixed spacetime backgrounds}, should be {\em finite} and {\em calculable}.
\item (B) Quantum fluctuations of spacetime itself, alias {\em spacetime foam},
should
be {\em tamed} and lead to {\em calculable} physical effects.
\end{description}

Traditionally, most of our efforts in string theory have been concentrated on
problem (A).
Building upon the deep insight of, most notably, G. Veneziano, Y. Nambu, M.
Virasoro, A. Neveu, J. Schwarz, P. Ramond, J. Scherk,  A. Polyakov, M. Green
and others, it was finally proven in 1992 by S. Mandelstam \cite{SM}
that indeed string theory provides {\em finite} and {\em calculable} QG
corrections
to any process, thus resolving problem (A). The method is standard by
now \cite{13,14}: one
chooses
a suitable CFT on the world-sheet to represent the {\em fixed} spacetime
background
($g_{\mu\nu}(X)$) usually taken to be flat Minkowski spacetime
($g_{\mu\nu}=\eta_{\mu\nu}$)
{\em plus} all other physical fields (``backgrounds") $g_i(X)$. In other words,
the CFT has a
{\em critical} or {\em fixed point} $g^*_i$, such that $\beta_i(g^*_j)=0$
represent successfully the E.O.M. of all the physical fields $g_i(X)$ in
target-space, according to Eq.~(\ref{24}).
In such a case we say that we got a {\em critical string vacuum}, represented
by the
corresponding CFT. Higher-order QG corrections correspond to higher-genus
effects, and
generalizations of Eq.~(\ref{25}) are available in such a case, providing {\em
finite} and
{\em calculable} results, as advertised above. While all these developments are
pretty
remarkable, unfortunately they do not address problem (B), \ie, the menace of
the
spacetime foam. In such a case, we need to take into account quantum
transitions between
{\em different} critical string vacua, and thus, at least ``{\em temporarily}"
we have to ``live" away from the {\em critical} or {\em fixed} point, and
consequently we have to generalize our notions of critical string theory
to {\em non-critical string theory}. Non-critical string theory, introduced
in 1988 by I. Antoniadis, C. Bachas, J. Ellis, and myself \cite{15}, provided
the
{\em first exact} string solution(s) (``vacuum(a)") corresponding to an
{\em expanding Universe}, in sharp contrast with the ``{\em static}"
background solutions of the critical string. Since non-critical string theory
\cite{15,16} is the {\em epicenter} of my discussions, it is useful to analyze
its structure
in some detail. In any closed string theory, as mentioned before, there is
{\em unavoidably} the massless graviton multiplet $\tilde g_{\mu\nu}$,
consisting of the spin-2
graviton $g_{\mu\nu}$ (symmetric and traceless part of $\tilde g_{\mu\nu}$),
the spin-0 {\em dilaton} $\Phi$ (containing the trace of $\tilde g_{\mu\nu}$),
and the {\em antisymmetric tensor} (of rank two) $B_{\mu\nu}$ (containing the
antisymmetric
part of $\tilde g_{\mu\nu}$), which only in four dimensions corresponds to a
pseudo-scalar $b$, called sometimes  the ``axion". In a way
this is a ``universal multiplet" contained in any closed string theory, and
thus
model independent. We shouldn't be surprised by this ``{\em universality}"
because the graviton multiplet provides the spacetime ``background" where
all other physical (target-space) particles ``live". It is natural that we can
have several theories ``living" on the same spacetime background. Thus,
if we are interested in quantum fluctuations of spacetime itself, \ie,
spacetime
foam, {\em to first approximation} we can neglect all other fields, beyond the
graviton multiplet, ``putting" them into their corresponding ground state, and
conveniently forgetting about them. The 2-d world-sheet action (\ref{21})
becomes
then \cite{13}
\begin{eqnarray}
S_{\rm 2-d}&=&-{T\over2}\int d^2\sigma[\sqrt{h} h^{\alpha\beta} g_{\mu\nu}(X)
\partial_\alpha X^\mu\partial_\beta X^\nu+B_{\mu\nu}(X)\epsilon^{\alpha\beta}
\partial_\alpha X^\mu\partial_\beta X^\nu]\nonumber\\
&&-{1\over2}\int d^2\sigma\sqrt{h}\Phi(X)R^{(2)}\qquad\qquad,
\label{26}
\end{eqnarray}
with $R^{(2)}$ the 2-d scalar curvature.
It is worth emphasizing some subtle difference between the origin of the
$(g_{\mu\nu},B_{\mu\nu})$ terms and the $\Phi$ term. As is indicated in
(\ref{26}), the $(g_{\mu\nu},B_{\mu\nu})$ terms are, for dimensional
reasons, proportional to the string tension $T$, while this is not the
case for the $\Phi$-(dilaton) term. Since ${1\over T}(\propto\alpha')$
acts as a world-sheet perturbation expansion parameter, it becomes apparent
that, if the $(g_{\mu\nu},B_{\mu\nu})$ terms appear at the ``classical
level", the dilaton term appears at the next order, as an ${\cal O}(\alpha')$
``quantum correction"! The first, rather obvious exact ``solution" to
(\ref{26})
is the ``Minkowski background":
\begin{equation}
g_{\mu\nu}(X)=\eta_{\mu\nu};\quad \Phi=\Phi_0,\quad B_{\mu\nu}(X)=0
\label{27}
\end{equation}
with the condition for {\em conformal invariance} ($\beta_\Phi=0$) implying
\begin{equation}
D+c_I=26
\label{28}
\end{equation}
where $D$ is the target space dimension and $c_I(\ge0)$ is the ``central
charge"
of the internal conformal system that, in principle, ``complements" the
$X^\mu$
fields on the world-sheet, so that the 2-d world-sheet QFT is
``conformal-anomaly"
free. Clearly there is a $D_{\rm max}=26$, the famous 26 dimensions that
the (bosonic) string has to ``live" in. It should be observed that the
``backgrounds" (\ref{27}) are {\em static}, \ie, no explicit time ($X^0$)
evolution!
Actually, this static ``solution" corresponds to the {\em critical string
vacuum}.
Clearly the critical string vacuum solution is the string analogue of the
usual
``Minkowski" solution of the {\em conventional point-like QFT}, which is the
infrared limit of the critical string. As such, the critical string inherits a
``small
problem" of conventional point-like QFT, mentioned in the previous sections,
notably the lack of an {\em arrow} of time which becomes endemic to critical
strings.
Actually, in a
way it may be even worse because it is possible to formulate {\em
consistently}
critical strings without a time-variable at all, \ie, on the light-cone gauge
with
$D=D_{\rm transverse}=24$. In such a case the physical meaning of $X^0$ is
similar to the physical meaning of the longitudinal photons in
quantum electrodynamics, not much!

Fortunately, there is another rather simple, non-trivial, {\em exact} solution
to
(\ref{26}) describing a {\em non-critical string vacuum} \cite{15}
\begin{equation}
g_{\mu\nu}(X)=\eta_{\mu\nu};\quad \Phi=-2QX^0+\Phi_0;\quad B_{\mu\nu}(X)=0
\label{29}
\end{equation}
with the condition for {\em conformal invariance} ($\beta_\Phi=0$) reading now
\begin{equation}
(D+c_I)-(12Q^2)=26
\label{30}
\end{equation}
where the ``background charge" $Q$ provides the ``central charge deficit"
$\delta c=12Q^2\ge0$. In such a case there is no $D_{\rm max}$ and the
(bosonic)
string may ``live" in {\em any} dimension ($\ge26$)! What really happens is
that by turning the dilaton $\Phi$ to be a {\em time-dependent} ($X^0$)
field,
we get an {\em extra} $Q$-dependent contribution to the $\beta_\Phi=0$
condition,
thus shifting the burden from $(D+c_I)$ to $Q$, such that (\ref{30}) is
always satisfied for {\em any value of $(D+c_I)(\ge26)$}! It is amusing to
notice
that {\em if and only if} $D=4$ there is another, closely related non-critical
string
vacuum (solution) \cite{15}, which ``activates" $B_{\mu\nu}$ by giving it also
a time
dependence,
and corresponds to a well-known CFT on the world-sheet, namely the Wess-Zumino
(WZ) model of an $SO(3)$ (or $SU(2)$) group manifold. The uniqueness of the
WZ
solution for $D=4$ follows from the simple observation that only for $D=4$,
\ie,
three space dimensions, the corresponding ``maximal symmetric space" ($S_3$)
is a group manifold ($SO(3)$)! We have been able to prove \cite{15} that the
``linear
dilaton" solution (\ref{29}), amended with the (WZ) solution if $D=4$, is the
{\em unique}
solution of the 1-loop (2-d) $\beta$-functions with $\delta c\not=0$ in the
following
sense \cite{15}:
\begin{description}
\item (i) Every possible solution approaches asymptotically (large $X^0$)
the solution (\ref{29}).
\item (ii) Even if we include {\em moduli fields} of the CFT coupled to the
dilaton $\Phi$,
again all possible solutions, for large $X^0$, approach asymptotically
(\ref{29})
and constant values of the moduli fields \cite{17}.
\end{description}
The {\em universality} and {\em uniqueness} of the ``linear dilaton"
non-critical
string vacuum is rather striking! Actually, there is more to it. The {\em
effective}
target-space $D$-dimensional action corresponding to the general action
(\ref{26})
may be written as
\begin{equation}
S^D_{\rm eff}=\int d^D X\sqrt{-g}
e^{-\Phi}\{R^{(D)}+(D_\mu\Phi)^2-{1\over12}H^2
-{1\over3}\delta c+\cdots\}
\label{31}
\end{equation}
with $R^{(D)}$ the target-space scalar curvature, $H_{\mu\nu\rho}$ the field
strength
of the antisymmetric tensor $B_{\mu\nu}$, and $\delta c$ a generic ``central
charge
deficit". The form of the action (\ref{31}) suggest that $\delta c$ can be
thought of as
an {\em allowed} tree-level cosmological term, as well as dilaton potential. We
are
thus faced with a horrendous {\em fine-tuning problem}: there is a priori no
theoretical reason why $\delta c$ should be {\em ab initio} zero, in other
words, ``{\em
static}"
Minkowski spacetime and constant dilaton (\ie, {\em critical strings}) are
{\em very
particular} ``solutions" of the string equations of motion. It sounds much more
reasonable
to start {\em generically} with $\delta c\not=0$ and then prove {\em
dynamically} that
we are ``relaxing" towards the $\delta c=0$ Minkowski ``solution"
asymptotically. Here
``relaxing" refers to the dependence of at least some of the physical fields on
some
dynamical variable, like for example the ``common sense" time, \ie, time {\em
with an
arrow}, as the cosmic time in an expanding Universe. It is heart-warming that
the
non-trivial solution (\ref{29}) provides {\em automatically} and {\em
generically}
$\delta c\not=0$, and a dependence of the dilaton $\Phi$-field on time $X^0$,
even
if at this moment $X^0$ {\em looks like} arrow-less Einstein time. But it is
not
so!
The ``physical metric", with a correctly normalized Einstein action is
\cite{15}
\begin{equation}
G_{\mu\nu}=e^{Q\Phi}g_{\mu\nu}=e^{Q\Phi}\eta_{\mu\nu}
\label{32}
\end{equation}
thus implying a Robertson-Walker line element
\begin{equation}
ds^2=c^2dt_c^2-t_c^2d\vec x^2
\label{33}
\end{equation}
with the cosmic time $t_c\propto e^{QX^0}/Q$, in other words a {\em linearly
expanding
Universe} \cite{15}!
It is highly remarkable that an {\em exact}, {\em universal}, and {\em unique}
{\em time-dependent} string solution exists, describing a {\em non-critical
string
vacuum}, with $\delta c\not=0$ and a {\em time-dependent dilaton} that
corresponds
to an {\em expanding Universe} \cite{15}. It seems that we have all the
prerequisites in
place
for a ``quantum relaxation" mechanism towards $\delta c=0$ (critical string
vacuum).

There is another, very interesting reason indicating that the ``linear
dilaton"
solution (\ref{29}) is really special: {\em the quantum origin of time}
\cite{18}.
Usually, when
one uses the classical (genus zero) 2-d action (\ref{26}), one assumes that
the
two-dimensional metric $h_{\alpha\beta}$ can always take the form
$h_{\alpha\beta}
=e^\phi\hat h_{\alpha\beta}$, with $\phi$ the conformal factor and $\hat
h_{\alpha\beta}$
some fixed metric. The conformal factor decouples from the classical 2-d
action,
although this decoupling is not automatic when quantum corrections are taken
into
account. The decoupling occurs {\em
also} at
the quantum level {\em if and only if} $\delta c=0$, otherwise, $\phi$ becomes
a dynamical degree of freedom,
the Liouville field.
Indeed, there is a quantum correction to the classical action (\ref{26}) given
by \cite{19}
\begin{equation}
S^{\rm quantum}_{\rm 2-d}=\int d^2\sigma\sqrt{\hat h}\{-\hat h^{\alpha\beta}
\partial_\alpha\phi\partial_\beta\phi+\sqrt{c_m-25}\,\phi R^{(2)}+\cdots\}
\label{34}
\end{equation}
where $c_m=3+c_I$ stands for ``matter" central charge. It is impressive that
quantum
corrections create a kinetic term for the conformal factor $\phi$, thus turning
it into
a new dynamical degree of freedom. It is even more impressive that this kinetic
term
has the wrong sign: {\em minus}! We just have to notice that (\ref{34}) can be
readily
absorbed in (\ref{26}) {\em if and only if} \cite{18}
\begin{equation}
\phi=X^0;\quad \Phi\sim\sqrt{c_m-25}\,\phi=\sqrt{c_m-25}\,X^0
\label{35}
\end{equation}
\ie, {\em if and only if} the string started in the ``linear-dilaton"
vacuum of (\ref{29})!
Furthermore, (\ref{34}) implies a conformal anomaly contribution of the
$\phi$-field:
$c_\phi=1-(c_m-25)$, or
\begin{equation}
c_m+c_\phi=26
\label{36}
\end{equation}
which is nothing else but the conformal anomaly cancellation condition
(\ref{30}),
reexpressed in terms of $c_m(\equiv3+c_I)$ and thus $c_m-25=\delta c=12Q^2$.
Remarkably, turning $\phi$ into a dynamical
degree
of freedom by moving to $c_m>25$, not only do we ``create" time by quantum
corrections
(\ref{34}), but {\em also} conformal invariance is imposed {\em dynamically}
(\ref{36}),
not put in by hand. This extra bonus was the missing link in identifying
string theory
with 2-d quantum gravity, and makes the whole picture much more logical and
aesthetically appealing. Using this 2-d quantum gravity language, one can
readily show \cite{15}
that the ``matter" part of the vertex operator ($e^{i\vec p\cdot\vec X}$) has
to be
``gravitationally dressed" by $e^{i(p^0+iQ)\phi}$ in order to render a
consistent
reparametrization invariant conformal field theory in 2 dimensions. Thus, we
explicitly
see how time ($\sim\phi$) is created and ``runs", while it is guaranteed that
the total
conformal weight is 1! It should be stressed that the dynamical appearance
of the Liouville field at the quantum level is in {\em perfect agreement}
with the origin of the dilaton ($\Phi$) term in the action (\ref{26}), which
as emphasized there, should be considered an ${\cal O}(\alpha')$ quantum
correction! The deep connection between the dilaton ($\Phi$), the Liouville
field ($\phi$) and time ($X^0$) is, of course, reflected in (\ref{35}).

It should be apparent by now that the ``time"-variable of the non-critical
string theory, as dynamically created by quantum corrections of the 2-d
world-sheet QFT, stands on a very different footing compared to the ordinary
space-dimensions ($\vec X$)! Actually, that is exactly what we were after
in order to resolve the {\em arrow} of time problem: space and time are indeed
different! As long as $\delta c=c_m-25=12Q^2\not=0$, we do live in an
``expanding Universe", with an {\em arrow} of time, and we do
asymptotically ``reach" a critical string vacuum (Minkowski
background), where $\delta c=0$ and where (\ref{34}) just provides us with
an Einstein-time of the proper {\em negative} signature! This way of looking at
things provides a very satisfactory picture: the ``idealized" Minkowski
background of special relativity, with all its Einstein-time consequences, is
only {\em asymptotically reachable} and thus {\em approximately true}, although
its phenomenal success of the last 90 years strongly suggests that we
are not very far from it! In other words, time has an {\em arrow} as long as
we are away from the critical string vacuum, \ie, away from criticality,
\ie, {\em out of equilibrium} (non-critical string vacuum), while it becomes
Einsteinian at {\em equilibrium} (critical string vacuum), in full accordance
with our standard notions of non-equilibrium quantum statistical mechanics,
that we now turn our attention to.

\section{Interlude (II) -- Friction Time}
It is very encouraging that in the case of non-critical string theory the
notion of {\em out-of-equilibrium} emerges because, as we discussed in
Section~3 (Interlude (I)), such notion may be required for a
``complete" understanding of Quantum Physics along the lines of (\ref{16}).
{\bf N}on-{\bf E}quilibrium {\bf Q}uantum {\bf S}tatistical {\bf S}ystems
[NEQSS] involve generically an {\em observed} ``open" (sub)system in contact
with an {\em unobserved} reservoir. The main characteristics of the {\em
observed (sub)system} relevant to our discussion here are: (i) the existence
of a ``microscopic" arrow of time related to (ii) an increase in the entropy,
$\Delta S_{\rm entr}>0$, in accordance with the Second Thermodynamic Law,
due to (iii) dissipation, which is also responsible for (iv) the ``collapse"
of the wave function: $\Psi_{\rm(sub)system}{\buildrel \rm collapses\over
\longrightarrow}\sum_i |c_i|^2|\Psi_i|^2$. Of course, this is exactly what we
are
after, as discussed in section 3. As indicated there, identifying the {\em
observed} ``open" (sub)system with the {\em low-energy}, {\em propagating}
particles and the {\em unobserved} reservoir with the {\em microscopic
event horizons} (unobservable even in principle) of the spacetime foam, may
get us somewhere. But, what are the conditions so that the wonderful properties
(i)-(iv) are realizable in practice? Thanks to the work of Misra and Prigogine
(MP) \cite{20} and their collaborators, we know the answer. In the MP theory
\cite{20} one assumes
the existence of an {\em non-unitary} $\Lambda$-transformation such that
\begin{equation}
\tilde\rho=\Lambda\rho;\quad
\partial_t\tilde\rho=\Phi(L)\tilde\rho
\label{37}
\end{equation}
where $\tilde\rho$ refers to the phase-space density matrix
describing the ``open" (sub)system, and $\Phi(L)\equiv\Lambda^{-1}L\Lambda$
is the Liouvillean of the (sub)system, while $\partial_t\rho=L\rho$ refers
to the corresponding quantities of the {\em whole}, {\em closed} system
((sub)system$\oplus$reservoir). Then, MP have proven \cite{20} that {\em under
certain
conditions} the dynamic ``open" (sub)system ($\tilde\rho$) {\em admits} an
internal time variable with an ``arrow", related to the increase of the ``open"
(sub)system entropy
\begin{equation}
\widetilde S=-{\rm Tr}\tilde\rho\ln\tilde\rho;\quad
\partial_t\widetilde S\ge0.
\label{38}
\end{equation}
They have also pointed out the strong bond that exists between their notion of
{\em time-irreversibility} and {\em nonlocality}, a kind of {\em inverse}
``butterfly effect". This very last point is of immense importance for us,
because
as we discussed in section 3, and will be the focus of our attention a bit
later, our rather restricted ability to ``measure" things {\em only} locally,
prevents us of taking into account the effects of extended, nonlocal (in
spacetime), solitonic string states, characteristically present in spacetime
foam, thus {\em necessarily} rendering our {\em ``idealized",
``local"} system, {\em open}! Let us consider \cite{3} the simplest possible
``open"
system, the {\em dissipative (\eg, friction) motion} of a single particle,
as described by the {\em open-version} of  Lagrange equations, involving
non-conservative forces (${\cal F}_i$),
\begin{eqnarray}
{\rm [2nd-order\ formalism]:}\quad &&\!\!\!\!\!\!{d\over dt}{\partial
L\over\partial \dot q_i}-{\partial L\over\partial q_i}={\cal F}_i(t,q_i,\dot
q_i)
\label{39}\\
{\rm or\ equivalently}\qquad\qquad\qquad&&\nonumber\\
{\rm [1st-order\ formalism]:}\quad \dot q_i&=&{\partial H\over\partial
p_i}\nonumber\\
\dot p_i&=&-{\partial H\over \partial q_i}+F_i\ ;\quad {\cal F}_i(t,q_i,\dot
q_i)=F_i(t,q_i,p_i),\nonumber\\
\label{40}
\end{eqnarray}
where as usual, $L$ and $H$ refer to the Lagrangian and Hamiltonian functions,
and the dots indicate time derivatives. Let us extend our study to the
statistical evolution of the phase-space density function $\rho=\rho(q,p,t)$,
dropping from now own the twidle over $\rho$ for simplicity. The generalized
Liouville equation for ``open" systems reads in our case \cite{3}
\begin{equation}
{\partial\rho\over\partial t}+\{\rho,H\}+F_i{\partial\rho\over\partial p_i}
+\rho{\partial F_i\over \partial p_i}=0
\label{41}
\end{equation}
with $\{,\}$ indicating the usual Poisson brackets. As is well known, we are
facing here a troublesome problem: while the ``physical energy" $E$ is not
conserved ($\dot E\not=0$) because of the  existence of
dissipative (\eg, friction) environmental effects that ``open" the system,
$H$, the generator of time-translations, is conserved $\dot H=\{H,H\}=0$.
In other words, we don't know anymore how $E$ relates to $H$! Fortunately,
there is a whole new branch of NEQSS that has been developed \cite{21}, under
the
name of {\em Lie-Admissible Algebras} (L-A-A), suitable to provide a
satisfactory answer. One has to generalize the notion of Poisson bracket
$\{,\}$, or in the quantum case, of the Lie product $[,]$, to {\em
symbolically}
$\{\{,\}\}$ and $(,)$ respectively. The new ``products" satisfy linearity,
generalized Jacobi indentities, and most importantly, in their quantum version,
the {\em Lie-Admissivity} property \cite{21}
\begin{equation}
(A,B)-(B,A)=2[A,B]
\label{42}
\end{equation}
which shows immediately that while $[H,H]$ is zero, $(H,H)$ {\em does not}
have to vanish! In such a case we get for any dynamical quantity $A$
\begin{equation}
\dot A=\{A,H\}+{\partial A\over\partial p_i}G_{ij}{\partial H\over\partial p_j}
\equiv\{\{A,H\}\}
\label{43}
\end{equation}
with $G_{ij}\equiv \delta_{ij}(F_i/({\partial H\over\partial p_i}))$, assuming
of course $\partial H/\partial p_i=\dot q_i\not=0$. For this simple dissipative
statistical mechanical system, the MP function $\Phi$ defined in (\ref{37})
reads \cite{3}
\begin{equation}
\Phi=i\sum_{i=1}^N\left(
{\partial H\over\partial q_i}{\partial\over\partial p_i}-
{\partial H\over\partial p_i}{\partial\over\partial q_i}\right)
+\sum_{i,j=1}^N{\partial H\over\partial p_i}G_{ij}{\partial\over\partial p_j}
\label{44}
\end{equation}
or equivalently (\ref{41}) reads
\begin{equation}
\partial_t\rho=-\{\rho,H\}+\sum_{i,j=1}^N\dot q_iG_{ij}{\partial
\rho\over\partial
p_j}.
\label{45}
\end{equation}
Remarkably enough, Constantopoulos and Ktorides (CK) \cite{22} have proven that
{\em all} the
MP \cite{20} conditions for time irreversibility etc, are satisfied, {\em
if and
only if}, $G_{ij}$ is {\em real} and {\em symmetric}, \ie\
$G_{ij}=G^*_{ij}=G_{ji}$.
In other words, we are after a {\em Lie-Admissible} NEQSS with {\em real}
and {\em symmetric} $G_{ij}\equiv\delta_{ij}(F_i/({\partial H\over\partial
p_i}))$. Well, we will show next that we got one such NEQSS: non-critical
string theory,
endorsed with the Zamolodchikov metric $G_{ij}$ \cite{23}, which happens to be
{\em real}
and {\em symmetric} for {\em unitary} 2-d CFT, as is always the case in
string theory!!!

\section{Back to the Present}
During the last three years, J. Ellis, N. Mavromatos, and myself have
utilized non-critical string theory as the right framework to address all
presently cumbersome problems analyzed in the previous sections. Our strategy
is the following \cite{1,2,3}: use the most general formalism of  2-d
world-sheet {\em unitary} and {\em renormalizable} QFT, generically
refered to as 2-d world-sheet $\sigma$-models, {\em without} the restriction
of being {\em necessarily} conformal field theories. A qualification is needed
here right away: as we discussed towards the end of section 4 (around
Eqs.~(\ref{34})-(\ref{36})), the ``conformal anomaly" condition (\ref{36})
is {\em always} satisfied, thanks to the cerberian behavior of the ``activated"
Liouville mode, which turns conformal invariance into a ``derived", dynamical
property. What we are discussing here is the rest of the system, \ie, {\em not}
including the Liouville mode, which is used as a dynamical evolution
parameter -- time -- as (\ref{35}) indicates. A very rough analogy would be
to allow a small ball to roll  between two points A
and B lying on the rim of a hemisphere, starting at rest, say, at A and
finishing at B assuming the dynamical energy relation $V_A=V_B$. We know that
total energy conservation implies that for every point on the ball trajectory,
the sum of the dynamical energy and the kinetic energy $T$ is constant, \ie,
{\em always}: $V+T=V_A=V_B$. {\em Nevertheless}, we learn a lot about the
ball's motion by concentrating our studies on the dependence of $V$ with
time, \eg, decreases some of the time and increases the rest of it, etc.
In our string case, the ``magic" number of 26 seems to be always conserved
(corresponding to the total energy of the ball), while $c_m$ and $c_\phi$
may change appropriately (corresponding to $V$ and $T$ respectively). Thus
we have the right to concentrate on studies of the ever changing ``matter"
part, $c_m$, and get invaluable information about the system out of it. The
deep question of why  $c_m$ {\em starts changing at all}, corresponding to
the small ``push" we have to give to our ball at A to start rolling, will
be addressed in due time!

\subsection{The formalism}
Let us start \cite{2,3} with a 2-d CFT represented on the world-sheet by the
action
$S_0(r)$, where $\{r\}$ refers to the ``matter" fields (\eg, $X^\mu:\
\mu=1,\ldots,D$) spanning a $D$-dimensional target-space manifold of
{\em Euclidean signature}, \ie, there is no time-variable! As usual, let us
perturb our system, or more properly in string theory, let us deform it, by
``turning on" some string background $g(r)$ represented on the world-sheet
by the (1,1) (in order to satisfy global scale invariance), but not
necessarily {\em exactly marginal}, vertex operator $V_g(r)$. Exactly
marginal operators, in technical jargon, means that their Operator Product
Expansion (OPE) $V_g\otimes V_g$
does not contain a term of the form $C_{ggg}V_g$, \ie, $C_{ggg}=0$. In
plain language, it means that such deformations never get the system out of
the {\em critical line}. In critical string theory, only {\em exactly marginal}
vertex operators appear/are allowed, because there is no dynamical Liouville
mode ($c_\phi=0$), thus $c_m$ has to remain {\em always} constant!

Using our small ball analogy, this means that the ball may, {\em at most},
move on a flat surface (equipotential surface),  \ie, $V=$constant, or in the
string language, $c_m=$constant. Actually here lies the whole {\em clue}
of our approach: by deforming our original system (CFT), with {\em not} exactly
marginal vertex operators, we are allowing $c_m$ to change, by ``utilizing"
the Liouville field ($c_\phi\not=0$). Let us see how this works \cite{2,3}. The
existence
of a non-exactly marginal vertex operator, \ie, $C_{ggg}\not=0$ implies the
existence of an ``anomalous" scaling dimension $\alpha_g=-gC_{ggg}+{\cal
O}(g^2)$. Thus, global scale invariance, a rather {\em untouchable} fundamental
prerequisite, is in jeopardy, unless the Liouville mode comes to the rescue,
``dressing" appropriately our non-exactly marginal operator, so that it keeps
its (1,1) scaling dimensions {\em intact}! Indeed, to first order in $\alpha'$,
we get \cite{2,3}
\begin{equation}
\int d^2z gV_g(r)\to\int d^2z ge^{\alpha_g\phi}V_g(r)=\int d^2z gV_g(r)
-\int d^2z g^2 C_{ggg}V_g(r)\phi+\cdots
\label{46}
\end{equation}
indicating clearly that global scale invariance is guaranteed {\em if and
only if} $g(r)$ gets {\em renormalized}, implying a non-vanishing
$\beta$-function for $g$
\begin{equation}
g_R\equiv g-C_{ggg}\phi g^2\to \dot g_R={dg_R\over d\phi}\equiv \beta_g=
-C_{ggg}g^2_R+\cdots\not=0
\label{47}
\end{equation}
with $\phi(\sigma,\tau)$ the Liouville mode, apparently playing the role of
a {\em local renormalization group scale} on the world-sheet
\begin{equation}
\phi(\sigma,\tau)\equiv\ln\Lambda(\sigma,\tau).
\label{48}
\end{equation}
But, as we discussed in detail in section 4 (around Eqs.~(\ref{34}-\ref{36})),
$\phi$ {\em acquires dynamics} through integration over world-sheet covariant
metrics $h_{\alpha\beta}=e^\phi\hat h_{\alpha\beta}$, $\hat h_{\alpha\beta}=$
fixed, and becomes an extra target coordinate with, in the case of $c_m\ge25$,
negative metric, \ie, the time-coordinate, $X^0$ (\ref{35}).
Thus, by combining (\ref{35}) and (\ref{48}) we arrive at the highly
remarkable {\em dynamical} relation \cite{2,3}
\begin{equation}
\phi(\sigma,\tau)=X^0(\sigma,\tau)=\ln\Lambda(\sigma,\tau)
\label{49}
\end{equation}
which clearly indicates the {\em singularly} {\em double} role of the
{\em dynamical} Liouville field $\phi$, ``active" only in the non-critical
string theory, as the time variable, as well as the local renormalization
group (RG) dynamical scale that ``flows" or ``runs" along the RG trajectories
of the 2-d world-sheet $\sigma$-models. This is the ``microscopic" explanation
of the quantum origin of time discussed in section 4 and, as (\ref{47})
indicates, the Liouville ``dressing" makes the string background $g(r)$
be time-dependent, $g(r,\phi)=g(r,X^0)$. Since the time variable $X^0$
is of a very different origin (in fact of {\em quantum} origin) compared with
the {\em classical} $\vec X$-fields representing the space-coordinates denoted
above collectively as $\{r\}$, we should not be surprised that time ``flows"
irreversibly,
while space does not. It looks like that we are on the right track.

The above analysis can be easily generalized to an arbitrary number of string
backgrounds $\{g_i\}$, with corresponding vertex operators $\{V_{g_i}\}$ and
$\beta$-functions $\beta_i\equiv\dot g_i$. We shouldn't forget the double role
of the $g_i$'s: they represent physical fields of target-space (graviton,
photon,
electron, ...), as well as coupling-constants (or functions) of the 2-d
world-sheet
$\sigma$-model. Viewing them as coupling constants, we can consider them as
the coordinates spanning the manifold of all {\em unitary} and {\em
renormalizable}
2-d world-sheet $\sigma$-models $G\equiv\{g_i\}$, which is endorsed with a
Zamolodchikov ($Z$) metric \cite{23}:
\begin{equation}
G_{ij}(g)=2|z|^4\VEV{V_{g_i}(z)V_{g_j}(0)}
\label{50}
\end{equation}
which is {\em real} and {\em symmetric}, due to the unitarity of the
$\sigma$-models
on the world-sheet. Furthermore, there is a function $C(g)$ defined on $G$,
such
that
\begin{equation}
{\delta C(g)\over \delta g_i}=G_{ij}\beta^j
\label{51}
\end{equation}
and thus
\begin{equation}
\partial_t C(g)=-\beta^i G_{ij} \beta^j\le0
\label{52}
\end{equation}
where the last step (``{\em positivity}" of $G_{ij}$) follows from the {\em
unitarity}
of the $\sigma$-models on the world-sheet, and for shorthand, $X^0$ has been
replaced by $t$. Zamolodchikov's $C$-theorem \cite{23} (\ref{51},\ref{52})
shows that at
fixed points ($\beta_i(g^*)=0$) $C(g)=$constant, and actually, it corresponds
to the ``central charge" of the CFT representing the $\sigma$-model at the
critical point $g^*$. It is also apparent from (\ref{52}) that {\em if} any of
the
$\beta_i$ is different from zero, then the system ``runs" irreversibly into
{\em smaller} values of central charge $C$! This {\em irreversible} ``flow"
along the RG trajectory (\ref{52}) is of particular importance for us for two
main
reasons. Firstly, the identification of time with the local RG scale
($\ln\Lambda(\sigma,\tau)$) that {\em irreversibly} ``flows" along the RG
trajectories,
provides us with a remarkably simple and aesthetically appealing solution to
the
{\em arrow} of time problem! The non-vanishing $\beta_i$ (see (\ref{47})) is
the
origin of the time-``flow" or {\em arrow} of time, and if it hits a
fixed
point
($\beta_g=0$), presumably corresponding to a critical string vacuum, time
stops
to ``flow", it becomes {\em arrow-less}, \ie, it becomes Einstein time!
Secondly,
if we
take into account the deep connection \cite{14,24} that exists between $S^{\rm
Target-space}_
{\rm eff}[g_i]$, where $g_i$ play the role of physical fields, and $C(g_i)$
\begin{equation}
S^{\rm Targ.sp.}_{\rm eff}[g_i]=C(g_i)
\label{53}
\end{equation}
it implies that the ``roll-over" towards smaller values of $C$ is nothing else
but
the usual ``roll-over" towards smaller values of the effective potential
$V_{\rm eff}$!
Maybe the example of a  small ``rolling" ball used in the beginning of the
section
as an analogy for a ``running" central charge was not ``off-the-wall" as it
may have
sounded. Combining now (\ref{51}) and (\ref{53}), we get \cite{2,3}
\begin{equation}
{\delta S^{\rm Targ-Sp.}_{\rm eff}[g_i]\over \delta g_i}={\delta C\over\delta
g_i}
=G_{ij}\beta^j
\label{54}
\end{equation}
which give us, recalling the fact that $\delta S_{\rm eff}/\delta g_i$ are just
the
Lagrange equations,
\begin{equation}
{d\over dt} {\partial {\cal L}(g_i,\dot g_i)\over\partial\dot g_i}
-{\partial{\cal L}\over\partial g_i}=G_{ij}\beta^j
\label{55}
\end{equation}
which is nothing else but (\ref{39}), describing the {\em dissipative motion}
of
particles, involving {\em non-conservative forces} ${\cal F}_i\equiv
G_{ij}\beta^j
=G_{ij}\dot g^j$, in full accord with the definition of $G_{ij}$ provided by
(\ref{43})!!!!
Considering next the {\em target-space density matrix} $\rho=\rho(g_i,p_i)$,
with
$p_i$ the conjugate momenta of $g_i$ ($p_i=\partial{\cal L}/\partial \dot
g_i$), we
get, thanks to the renormalizability of the world-sheet $\sigma$-model,
\[
{d\rho[g_i,p_i,t]\over d\ln\Lambda}={d\rho[g_i,p_i,t]\over
d\phi}={d\rho[g_i,p_i,t]\over dt}=0
\]
which yields
\begin{equation}
{\partial\rho\over\partial t}+\dot g_i{\partial\rho\over\partial g_i}+\dot
p_i{\partial\rho\over
\partial p_i}=0
\label{56}
\end{equation}
Using the equations of motion, in their 1st order or Hamiltonian form, similar
to (\ref{40}),
it is easy to see that (\ref{56}) is identical to (\ref{41}), but with
${\partial F_i/\partial
p_i}=0$. Thus, by following the steps between (\ref{41}) and (\ref{45}),
(\ref{56}) may
be cast in the form \cite{25}, similar to (\ref{45}),
\begin{eqnarray}
\partial_t\rho&=&-\{\rho,H\}+\sum_{i,j=1}^N\dot g_i
G_{ij}{\partial\rho\over\partial p_j}
\nonumber\\
&=&-\{\rho,H\}+\delta\H\rho
\label{57}
\end{eqnarray}
which is nothing else but the highly desirable master equation (ME) (\ref{9}),
but with
$\delta\H$ {\em explicitly prescribed} \cite{25}
\begin{equation}
\delta\H\rho=\sum_{i,j=1}^N\dot g_i G_{ij}{\partial\rho\over\partial p_j}
\label{58}
\end{equation}
a rather remarkable result. Before losing (in)sight, let us reiterate the
following:
in the framework of non-critical string theory, and {\em assuming that some
$\beta_{g_i}
\not=0$}, we have found that time is of quantum origin (\ref{48}), and as such
it has
an {\em arrow}, related (\ref{49}) to the irreversible RG flow (\ref{52}) on
the
world-sheet.
The Liouville equation (\ref{57}) for the target-space density matrix, takes
its
{\em Lie-Admissible-open-system form} (\ref{45}) but with $G_{ij}$ identifiable
with
the Zamolodchikov metric (\ref{50}), {\em real} and {\em symmetric}, thus
fulfilling the
Misra-Prigogine conditions \cite{20}, so that the wonderful properties (i)-(iv)
(discussed in the
beginning of section 5) are inherited by our system! The self-consistency of
the
whole picture is striking: the {\em irreversibility} of the RG flow,
identifiable with
{\em time-irreversibility}, is {\em due} to the ``{\em positivity}" of
$G_{ij}$, \ie,
{\em real} and {\em symmetric} $G_{ij}$, emerging from the {\em unitarity} of
the
2-d world-sheet $\sigma$-models, which is {\em exactly} the same condition
that
CK \cite{22} found, such that the MP conditions are met and
thus entailing the existence of microscopic time with an {\em arrow}!!!
Furthermore,
as we will discuss shortly, the increase of entropy (as demanded by the Second
Thermodynamic Law) is {\em automatic} thanks to the ``{\em positivity}" of the
$G_{ij}$!
{\em Amazing}. We will find it very useful in the following to encode the above
correspondence between non-critical strings and dissipative systems, in a
lexikon-like way
\begin{equation}
\begin{tabular}{lcl}
$\sigma$-models ``coupling constant"&$\longleftrightarrow$&
dissipative, Lie-Admissible\\
phase space $(g_i,p_i)$&& particle statistical system\\
&&phase space $(q_i,p_i)$\\
({\em Non})-criticality of strings&$\longleftrightarrow$&({\em
Non})-conservative forces\\
$(G_{ij}\beta^j\not=0)$&$\longleftrightarrow$&$({\cal F}_i\not=0)$\\
$C(g_i)$&$\longleftrightarrow$&$S_{\em eff} [q_i]$\\
{\bf E.O.M.}:  ${\delta C\over\delta g_i}=G_{ij}\beta^j$&$\longleftrightarrow$
&${\delta S_{\rm eff}\over\delta q_i}={\cal F}_i$
\end{tabular}
\end{equation}
Before we enjoy and discuss further the fruits of our labor, we have
to deal
with the rather {\em pressing} question of who provides us and how with the
{\em underlying central assumption} of all our formalism: $\beta_i\not=0$?

\subsection{The microscopic mechanism -- A new fundamental Procrustean
Principle}
In order to take into account the effects quantum fluctuations of spacetime,
\ie,
spacetime foam, we have to {\em allow} for quantum transitions between
different
critical string vacua (CFT's). Since each critical string vacuum (CFT)
corresponds
to a critical or fixed point of the $G$-space, where all $\beta_i(g)=0$, once
we decide to move
from one critical point to another, we will, in principle, {\em unavoidably} go
through
non-critical points of $G$-space, where at least some of the
$\beta_i(g)\not=0$!
Actually, all the formalism \cite{2,3} of non-critical string theory developed
just above
takes
into account exactly this point, and the renormalization group (RG)
trajectories
discussed above are just the trajectories that connect two fixed points, with
in
principle a lot of non-critical points in between. While some ``{\bf G}rand
{\bf O}ld
{\bf S}tring {\bf T}heorists" (GOSTs) may get a bit worried and upset about
allowing non-critical points ($\beta_i(g)\not=0$) into string theory, they
really
shouldn't! Let me explain. In the past, conformal invariance was fixed by
hand
by not allowing the Liouville field $\phi$ to get activated, and it had been
just there
disconnected and forgotten. We found in recent years, that this ``negligence"
of
the Liouville field was a very special case, and really unjustifiable. The
Liouville
field is ``activated" in any ``realistic" string solution (\ie, expanding
Universe) and
its central charge, automatically and naturally takes
care
of conformal invariance, \ie, dynamically establishes conformal invariance. A
much
more satisfactory state of affairs. In such a case, ``running" along the RG
trajectory
between two different critical or fixed points, {\em with $\phi$ as the RG
scale},
assures that the {\em whole} system (``matter" $\oplus$ Liouville field) is
conformally
invariant {\em at every point of the RG trajectory}. In our analogy with the
``rolling"
of the small ball, it corresponds to the conservation of the total energy
($V+T$) along
every point of the ball trajectory. On the other hand, the ``running"
corresponds
to a {\em dynamical} evolution of our ``matter" system that we are really
interested in,
\ie, not just ``playing around" a given fixed point, but moving all the way
from one
fixed point to another. In our analogy with the small ``rolling" ball example,
it
corresponds to leaving the ball ``run" its course, as it moves from one
``height"
to another, in principle significantly different. We, kind of, bring to {\em
completion}
the close analogy that exists between string theory, viewed as 2-d
world-sheet
$\sigma$-model and 2-d statistical mechanical system {\em at a critical or
fixed point},
by just extending the analogy beyond
the {\em critical point}! The {\em dynamically recovered} conformal invariance
unleashes our hands and allows us to treat string theory
as
a genuine statistical mechanical system. In such a case, we have to consider
{\em
not only} {\bf E}{\em xactly} {\bf M}{\em arginal} {\bf O}{\em perators} (EMO),
that {\em at most} can take
you along
a critical line, but also {\bf R}{\em elevant} {\bf O}{\em perators} (RO) that
``run" away from a
given
fixed point, and {\bf IRR}{\em elevant} {\bf O}{\em perators} (IRRO), that
``run" towards a fix
point.
In the small ``rolling" ball analogy, they correspond to pushing the ball away
from the inverted bottom,
or bottom, of the hemisphere respectively. Both relevant and irrelevant
operators
are characterized by $\beta(g_i)\not=0$, positive for RO and negative for IRRO,
while
as discussed previously, $\beta_i(g)=0$ for EMO. Thus in the case of
non-critical
strings, and by their close correspondence with a genuine statistical
mechanical
system, we have to allow the existence of IRR and R operators as possible
fluctuations
or deformations, in other words, we have to allow $\beta_i(g)\not=0$. Thus we
shouldn't
be surprised after finding all these wonderful analogies with (non)equilibrium
quantum
statistical mechanics, that the {\em desiderata} for understanding the
{\em arrow} of
time and the ``collapse" of the wavefunction may be met in non-critical string
theory.
The captivity, at the critical point, is over!

Let me describe now the microscopic mechanism that we think is at work,
providing
{\em effectively} non-vanishing $\beta$-functions ($\beta_i(g)\not=0$). I will
use the
two-column format, so that the one-to-one correspondence  between world-sheet
phenomena and target-space phenomena remains apparent during the whole
process. For the ``doubting Thomasses", in the section  8 I will discuss very
briefly,
since it has been discussed and reviewed {\em numerous times} by now
\cite{1,2,3}, a ``toy"
{\em 2-D (spacetime)} black-hole model, due to Witten \cite{26}, where {\em
all} the
forthcoming two-column steps have been worked out, by us, {\em explicitly} and
in {\em excruciating detail}.

\vbox{
\begin{flushleft}
\begin{tabular}{lcl}
{\bf WORLD-SHEET}&&{\bf TARGET-SPACE}\\
&&\\
Topological ``defects"&\llra&\parbox{3in}{spacetime foam, containing
microscopic event horizons due to
creation and annihilation of
microscopic (Planck size) black holes}\\
Need:&&\\
\parbox{3in}{$\left\{\begin{array}{ll}\rm ``Monopoles"\cite{27,28}\\\rm
``Antimonopoles\
at\ infinity"\end{array}\right.$}&\llra&
\parbox{3in}{black-hole \cite{28}$\left\{\begin{array}{ll}\rm horizon\\\rm
``singularity"\end{array}\right.$}\\
``Instantons" \cite{29,30}&\llra& black-hole quantum decay \cite{31,30}\\
&&\\
\parbox{3in}{existence of:\\
``monopoles"/``instantons" {\em imply}\\
the existence of:\\
``gauge symmetries"\\
Increase {\em usually} available symmetries}&\llra
&\parbox{3in}{Increase available\\ local symmetries}\\
&&\\
\parbox{3in}{\eg: usual Virasoro algebra:\\
$[L_m,L_n]=(m-n)L_{m+n}+\cdots$\\
is contained in the $W_\infty$ algebra \cite{32}\\
$[W^i_m,W^j_n]=[jm-in]W^{i+j-2}_{m+n}+\cdots$}&&
\parbox{3in}{\eg: Black holes with ``W-hair"\\
\cite{1,33}\\
(infinite dimensional,\\ commuting, conserved\\ charges)}\\
&&\\
\parbox{3in}{Close to the 'Hooft-Polyakov\\ monopole center, and in
{\em sharp contrast} with the would-be {\em Dirac-monopole singularity},
the corresponding order parameter(s) vanish ($\VEV{W}=0$), thus enlarging
the symmetry, \ie, we are getting the primordial Big Symmetry ($\Omega$) which
corresponds to a Topological Field Theory \cite{34},
involving ``surface" terms a la Chern-Simons or Wess-Zumino (WZ)}
&\llra&\parbox{3in}{Close to the would-be (spacetime) classical ``singularity"
the spacetime metric (order parameter) vanishes ($\VEV{g_{\mu\nu}}=0$), one of
many vanishing order parameters, thus getting us to the primordial
Big Symmetry ($\Omega^T$) which corresponds to a
Topological Field Theory \cite{26,35,36}}
\end{tabular}
\end{flushleft}
}

\vbox{
\begin{flushleft}
\begin{tabular}{lcl}
(...continued)&&\\
{\bf WORLD-SHEET}&&{\bf TARGET-SPACE}\\
&&\\
\parbox{3in}{Primordial Big Symmetry ($\Omega$)\\
``{\em breaks down}" \cite{36} (via ``instantons")
to $W_\infty$ symmetry
(thus a ``relic" of topological symmetries)}&\llra
&\parbox{3in}{Primordial big symmetry ($\Omega^T$) ``{\em breaks down}"
\cite{36,30,3} (via $\VEV{g_{\mu\nu}}\not=0,\ldots$)
to ({\em spacetime})$\oplus$topological-states
(``relics" of the topological phase)}\\
&&\\
\parbox{3in}{``Area-preserving diffeomorphisms \cite{33} of $G$-manifold
phase-space"}
&\llra&\parbox{3in}{{\em Quantum coherence} \cite{33,1}\\ ($\delta\H=0$)!}\\
&&\\
\parbox{3in}{any available deformation operator is Exactly Marginal Operator
(EMO)}
&\\
&&\\
{\bf BUT}&\\
&&\\
\parbox{3in}{$V_g=\left((V_g)_{\rm ``light"}\oplus\sum_i``W_i"\right)=$\\
(due to $W_\infty$ symmetry$\uparrow$)\\
= Exactly Marginal Operator (EMO)}
&\llra&\parbox{3in}{$g=(g)_{\rm local}\oplus\sum_i ``g_{W_i}"$\\
\hspace*{3cm}\parbox{1.5in}{$\uparrow$ {\em nonlocal},\\ ``{\em
topological}",\\ {\em extended \\ states}}}\\
\hspace*{2cm}{\em thus}&&\\
&&\\
\parbox{3in}{$(V_g)_{\rm ``light"}$ cannot be EMO}&\llra&
\parbox{3in}{The ``$g_{W_i}$", because of their {\em delocalized nature}, they
neither appear as well-defined asymptotic states nor can they be integrated out
in a {\em local} path-integral formalism, thus defying their detection in
{\em local} scattering experiments \cite{25,2,3}}\\
&&\\
\parbox{3in}{\hspace*{2cm}{\em and}\\
 ``effectively" becomes\\
\hspace*{2cm}{\em a}}&& \parbox{3in}{\hspace*{2cm}{\em thus}\\ ``effectively"
$g$ becomes}\\
&&\\
Relevant operator!&\llra&$(g)_{\rm local}$\end{tabular}
\end{flushleft}
}
\newpage
\begin{flushleft}
\begin{tabular}{lcl}
(...continued)&&\\
{\bf WORLD-SHEET}&&{\bf TARGET-SPACE}\\
&&\\
implying \cite{25,2,3}&&\parbox{3in}{\em Unavoidably implying spontaneous\\
truncation}\\
&&\\
\parbox{3in}{$\beta_g=(\beta_g)_{\rm ``light"}+\sum_i\beta_{``W_i"}=0$\\
$\Longrightarrow(\beta_g)_{\rm ``light}\not=0$}&\llra&
\parbox{3in}{
{\Large$\rho_{(g)_{\rm local}+``g_{W_i}"}\to\tilde\rho_{(g)_{\rm local}}$}\\
(\ system\ \ \ \ \ \ \ \ \ \ \ \ (sub)system)\\
$(\delta\H)_{\rm ``light"}\not=0$\\
``effectively" quantum\\ coherence loss for\\ local/``light" subsystem
\cite{25,2,3}}\\
&&\\
\end{tabular}
\hrule
\end{flushleft}

Some explanations are due. It shouldn't be surprising that topological
non-trivial ``solutions" of target-space, like black-holes, correspond to
topological nontrivial structure on the world-sheet, including
``monopoles" and ``instantons". What else could it be? Furthermore, there is a
one-to-one correspondence \cite{28} between the
``monopole" charge and the black-hole mass. Actually, ``instantons" may cause
transitions between ``monopoles" of different charges, corresponding to
higher-genus (quantum level) corrections to black-hole masses, thus triggering
{\em quantum decay} \cite{2,3,30,31} of the black hole. The accommodation of
topological non-trivial structure on the world-sheet, needs  gauge symmetry,
which forces the Virasoro algebra to become a $W_\infty$ algebra, which
through the Evans-Giannakis-Ovrut (EGO) mechanism \cite{37}, translates to
infinite dimensional,
commuting, conserved charges on target-space, that endorse the black-hole.
The pipe-dreaming of section 3 becomes a reality: non-critical strings provide
non-trivial, infinite dimensional symmetries, whose corresponding conserved
charges characterize the identity of the black hole. Thus, enabling the B.H.
to absorb all information, in a coherent way, and then, through normal {\em
quantum decay} \cite{31}, and not through {\em thermal Hawking radiation}, to
spit it out coherently, through a very {\em constrained} {\em decay-cascade}
\cite{38}, due to the severe
obligation of the decay process to respect a whole set of selection rules.
The {\em common} origin of the ``melting" of the singularity (``Dirac-like"
in the ``monopole" case and ``space-like" in the B.H. case), due to the
enhancement of the infinite symmetries as we approach the ``singularity" and
thus unavoidably rendering {\em both} world-sheet and target-space actions
{\em topological}, is fascinating. Normal QFT, 2-d (world-sheet) or
$D$-dimensional (target-space), based on Riemannian metrics, cannot sustain
such an incredible amount of infinite symmetry and they disolve to
topological field theories! The infinite symmetries that characterize string
theories, away from the ``classical" singularity, \ie, in the ``spacetime"
phase, are nothing else but a subgroup of the ``infinite" symmetries in the
topological phase! Thus, we should not be surprised that in the ``spacetime"
phase we may encounter \cite{33} nonlocal, topological, extended states, \ie,
states with {\em definite, discrete} $E$ and $\vec p$ (extendable over {\em all
space} and {\em all time}, thanks to the uncertainty principle!), the ``relics"
 \cite{36} of spontaneously broken infinite symmetries, something like the
Goldstone bosons of spontaneously broken global symmetries. It is exactly the
very existence \cite{33} of these delocalized states which, since they {\em
cannot be accounted} by a {\em local} observer, renders the {\em local} system
{\em unavoidably, open} \cite{2,3,25}! Undoubtedly, we have discovered
\cite{2,3,25}
here a very interesting, new and of {\em pure stringy origin} phenomenon, which
is the source of the conceptual changes that may be needed at the foundations
of Quantum Physics and Relativity. It seems very likely that we are uncovering
\cite{2,3,25} a {\em new fundamental principle}, beyond the
{\em wave-particle duality} (or {\em uncertainty principle}) of Quantum
Mechanics and the {\em Relativity principle}, which in a way completes the
set of our big sacred principles. It may be expressed in a very general,
{\em abstract} way as follows.

{\em In our ``low-energy" world our available ``tools", made of the lowest
lying, {\em propagating} string modes, are localized, thus the available
``measurements"/ ``experiments" are necessarily localized. As such, we
cannot take into account the {\bf complete} effects of the topological,
extended (over all space-time) states, in continous interaction with the
low-energy propagating particles. Thus, a {\bf spontaneous truncation} occurs
in a natural, unavoidable and {\bf dynamical} way, that renders a local system
wide open. We ``measure" only what fits in our ``apparatus", and thus let me
call it, appropriately, the {\bf Procrustean Principle}.}

Technically, such a procedure would amount to integrating out the topological
modes in a String Field Theory path-integral. Since a second quantized
String Field Theory is not yet at hand, a first quantized $\sigma$-model
approach \cite{13} can be adopted instead, in which such an integration
would appear as turning on a ``{\em spontaneously truncated}" background
of propagating light string modes only \cite{2,3}.

After this rather detailed discussion of the {\em formalism}, and of the
{\em microscopic mechanism} at work in non-critical string theory, responsible
for the ``openess" of the ``low-energy" world available to us, it is
appropriate to discuss some of the rather dramatic consequences that may lie
in our

\section{Future}
It became apparent in the previous section that non-critical string theory
may furnish the appropriate framework to address and correlate all the ``deep"
problems that have plagued us for almost a century now: ``arrow" of time,
``collapse" of the wavefunction, quantum gravitational dynamics, including
black-holes. As it was conjectured in section 3, the resolution of these
problems seems to cast down the {\em shibboleths} of conventional physics,
but in a very intriguing way.

\subsection{Density matrix mechanics}
The central issue of non-critical string theory, viewed as a
{\bf N}on-{\bf E}quilibrium-{\bf Q}uantum-{\bf S}tatistical-{\bf S}ystem
(NEQSS), is the derivation
(see (\ref{57}),(\ref{58})) of the Liouville equation for an ``open" system,
which becomes\\
{\bf A} {\bf M}aster {\bf E}quatio{\bf N} (AMEN) \cite{25}:
\begin{eqnarray}
\partial_t\rho&=&-\{\rho,H\}+
\left({1\over M_{Pl}}\right)G_{ij}\beta^j{\partial\rho\over\partial p_i}=
i[\rho,H]+\left({1\over M_{Pl}}\right)iG_{ij}[g_i,\rho]\beta^j,\nonumber\\
{\rm implying}&&\nonumber\\
\delta\H\rho&=&\left({1\over
M_{Pl}}\right)G_{ij}\beta^j{\partial\rho\over\partial p_i}=
\left({1\over M_{Pl}}\right)iG_{ij}[g_i,\rho]\beta^j
\label{60}
\end{eqnarray}
where the second term holds in the quantum formulation. As I have indicated
explicitly, the extra term in the standard Liouville equation is at least of
${\cal O}(1/M_{Pl})$ with respect to the standard term $[\rho,H]$. Since the
origin of this extra term is due to the mixing (triggered by the
$W_\infty$-symmetries) between the ``low-energy world" ($E$) particles and the
massive (${\cal O}(M_{Pl})$), higher level, topological string modes, we
naturally expect $(\beta^j)_{``light"}$, a concise ``measure" of
non-marginality, to be proportional to this mixing, \ie, $(\beta^j)_{``light"}
\approx{\cal O}((E/M_{Pl})^n)$, with $n$ some small positive integer. A ``tiny"
extra term indeed! But as Maxwell taught us (see section 1, after (\ref{4}))
``Beware of {\em tiny} terms bearing changes"! Indeed, this time too the
consequences are rather far-reaching. To start with, even the mere presence
of this ``tiny" $\delta\H$ term, eroding the wave-behavior of matter, entails
the abandoning of
{\em even the notion} of the wavefunction $\Psi$ and replacing it
by the density matrix $\rho$. Consequently, we have to \cite{2,3,25} abandon
the notion of
the $S$-matrix: $|\alpha,+\infty\rangle=S|\alpha,-\infty\rangle$ and replace
it by the $\S$-matrix: $\rho_+=\S\rho_-$, with $\S=SS^\dagger+\delta\S$. While
these remarks sound suspiciously similar to the remarks made in section 2
(around (\ref{8}),(\ref{9})), there is a {\em big difference}. There we didn't
have any consistent quantum gravity dynamics, we didn't know if our
conclusions were true or artifacts of semiclassical approximations, and we
didn't have any slight clue of how to calculate $\delta\H$, etc. Here we
have a consistent quantum gravity framework, and a microscopic mechanism
that explains how, starting from a {\em consistent} and {\em complete}
(non-critical) string theory at the quantum level, we end up {\em unavoidably}
and {\em necessarily} with an {\em effectively ``open"} non-equilibrium
quantum statistical system. As such, we can work out the explicit
form of $\delta\H$ or $\delta\S$ and thus we are able to provide new
rules for performing calculations. A very different picture from the murky
point-like QFT picture indeed! Furthermore, the specific structure/form of
the AMEN (\ref{60}) leads {\em immediately} to the following properties,
{\em independently} of the specific system under consideration \cite{2,3,25}:
\begin{description}
\item (I) {\em Conservation of Probability $P$}
\begin{equation}
\partial_t P=\int dp_i dg^i {\rm Tr}\left[{\partial\over\partial p_i}
(G_{ij}\beta^j\rho)\right]=0
\label{61}
\end{equation}
\item (II) {\em Conservation of Energy, on the average}
\begin{equation}
\partial_t\VEV{\VEV{E}}\equiv\partial_t\left[{\rm Tr}(\rho E)\right]
=\partial_t(p_i\beta^i)=0
\label{62}
\end{equation}
where the last equality is due to the {\em renormalizability of the
$\sigma$-model}. The latter
implies that any dependence on the renormalization
group scale in the $\beta^i$ functions
is implicit through the renormalized couplings.
Renormalizability replaces the time-translation
invariance of conventional target-space field theory.
\item (III) {\em Increase in Entropy}
\begin{equation}
\partial_t S\equiv\partial_t\left[-{\rm Tr}(\rho\ln\rho)\right]=
(\beta^iG_{ij}\beta^j)S\ge0
\label{63}
\end{equation}
where the last inequality follows from the {\em unitarity of the
$\sigma$-model}, as explained in section 6, see (\ref{50}).
\end{description}

These are three properties of fundamental importance for us, because
we are not talking about any random ``open" system, in which {\em generically}
energy conservation for the (sub)system is not guaranteed, but for a specific
``open" system that we claim describes the microcosmos and thus we cannot
afford energy non-conservation, and indeed we do get energy conservation!
Actually, it is intuitively easy to understand why. Our ``light" system
interacts with the massive topological models of {\em definite}, {\em discrete}
$E$, of the order of $M_{Pl}$, thus the ``light" system cannot
excite them and consequently, on the average, the $E_{\rm light}$ is conserved!
Similarly, the increase in entropy due to the
``{\em positivity}" of $G_{ij}$ is nothing else but the Zamolodchikov's
$C$-theorem \cite{23} condition for irreversible RG flow (see
(\ref{51},\ref{52})), \ie,
for irreversible time ``flow", as it should be in any NEQSS, like non-critical
strings, where the monotonic increase in entropy may be used to define the
``arrow" of time. It is highly remarkable that two fundamental properties
of world-sheet physics, {\rm renormalizability and unitarity} are responsible
for two fundamental properties of target-space physics, {\em conservation of
energy} and {\em monotonic increase of entropy} (implying {\em an arrow of
time}) respectively.

However, the renormalizability of the theory, which guarantees energy
conservation, does not guarantee the conservation of angular momentum
(see second Ref. in \cite{3}). Unlike string contributions to the increase
in entropy, which cannot cancel, the {\em apparent} non-conservation of
angular momentum may vanish in some backgrounds, though not in one cosmological
background that we have studied (see second Ref. in \cite{3}). Namely, in the
case of a maximally symmetric D-dimensional, non-static Universe with
$\partial_t R(t)\equiv -H(t)R(t)$ where $H(t)$ is a Hubble parameter, we find
(see second Ref. in \cite{3})
\begin{equation}
\partial_t\VEV{\VEV{J^{\alpha\beta}}}=
-{3D\over8\pi^2}H(t)R(t)\VEV{\VEV{J^{\alpha\beta}}}
\label{64p}
\end{equation}
showing a decrease of the average angular momentum in an expanding Universe.
This amounts to a {\em derivation} of {\em Mach's Principle}.

To discuss further some general properties of density matrix mechanics, I will
use again the two-column format employed in the last section, in order to
keep crystal clear the correspondence between world-sheet phenomena
and target space phenomena.

\begin{flushleft}
\begin{tabular}{lcl}
{\bf WORLD-SHEET}&&{\bf TARGET-SPACE}\\
&&\\
\parbox{3in}{``Monopoles"/``Instantons"}&\llra&\parbox{3in}{spacetime foam}\\
&&\\
\parbox{3in}{Trigger {\em Renormalization} ($\ln\Lambda(\sigma,\tau)$ or
$t$-dependence) of the $k$-parameter(s), \ie, the levels of the Kac-Moody
``currents" defining the CFT, expressible as $\theta_{\rm QCD}$-like
angles in the case of the Wess-Zumino type CFT, due to ``Instanton"
$({\rm IRR})\oplus(V_g)_{\rm light}({\rm R})$
deformations \cite{30,2,3}:
\[k{\buildrel t\to\infty\over\propto} e^{(\beta_g)gt}\to\infty,\]
 while
$k{\buildrel t\to0\over\longrightarrow} k_{\rm top.phase}$, a small finite
number, \eg, 2. In other words, a {\em monotonic increase} of $k$ with time
$t$, as we move away from the topological phase \cite{30,2,3}.}&\llra
&\parbox{3in}{Generates irreversible time-``flow" by getting the (sub)system
out of equilibrium, thus triggering {\em dynamically quantum relaxation} of all
the ``{\em constants}" \cite{2,3} of Nature towards their asymptotic (critical
string vacuum) values; $c,\hbar,l_{\rm string},\Lambda_C$ all become {\em
monotonically decreasing} functions of $k$, \ie, of $t$, as we move way from
the topological phase \cite{2,3}.}\\
&&\\
\parbox{3in}{The correlation functions $\VEV{V_1\cdots V_N}$ get
``non-perturbative" contributions (``valleys", ...) \cite{30}}&\llra
&\parbox{3in}{The scattering amplitudes\\ $A(1,2,\ldots,N)$ get non-standard
(\ie, beyond those included in (\ref{25})) contributions leading \cite{30} to
$\rho(t)=\S(t)\rho(0)$, with $\S(t)$ containing a factor $e^{-\gamma t}$, where
$\gamma$ represents generically the small anomalous dimension of the
deformation at hand. This extra exponential damping with time of the
$\S$-matrix leads, of course, to the {\em spontaneous collapse} of the ``wave
function" \cite{30,2,3}.}\\
&&\\
\end{tabular}
\hrule
\end{flushleft}

A few explanations are in order. Having convinced ourselves that
$\delta\H\not=0$,
\ie, that we are dealing with an ``open" system in the low-energy world, it is
only
natural that $\delta\S\not=0$ and thus non-factorization of the
$\S$,
triggers the {\em spontaneous collapse} of the wavefunction, irrespectively
of
the nitty-gritty microscopic details! Still, it is very interesting that the
topological
non-trivial structure on the world-sheet, including ``monopoles" and
``instantons",
contributes to the correlation functions and while this is nothing unusual at
the
world-sheet level, when these contributions are elevated to the target-space
level, they go far beyond \cite{30,2,3} usual QFT contributions. Furthermore,
the NEQSS
nature of the low-energy world makes it apparent that we are ``running"
towards
(but not yet there) the ``chosen" (critical string vacuum) fixed point. Thus,
the ``coupling constants" $g_i$ (our stringy backgrounds) evolve continuously
with time, and so do the associated world-sheet QFTs and thus their generic
parametrizations
(\ie, the $k$'s) become time-dependent \cite{30,2,3}. The case corresponding to
the
Wess-Zumino
CFTs is illuminating because $k_{\rm WZ}\to\theta_{\rm QCD}$-like angle,
and
we   know that ``instanton" effects {\em renormalize} $\theta_{\rm QCD}$!
Correspondingly, the ``constants" of Nature  that characterize a
given (critical string) vacuum, would depend on the parametrization of this
vacuum,
\ie, will be $k$-dependent and thus $t$-dependent \cite{2,3}.
Actually,
it is intuitively simple to determine their generic behavior with time. Close
to the
topological phase, \ie, $t\to0$, one expects the dominance of the topological,
extended states, thus the dominance of {\em nonlocal effects}, making it easy
to
communicate at {\em any distance}, thus suppresing the ``horizon". In other
words
allowing the velocity of light $c\to\infty$! As time goes by, the influence of
the
topological modes decreases and $c\to {\rm finite}$. The above discussion
shows
rather clearly that $c=c(t)$ is a {\em monotonically decreasing} function of
$t$.
In the same spirit, the closer we get to the topological phase, the easier it
is
for the (sub)system to become ``closed" and ``complete", \ie, standard quantum
mechanics at its full strength, with maximal $\hbar$! Again, $\hbar=\hbar(t)$
is a {\em monotonically decreasing} function of $t$, because getting away from
the topological phase the influence of the topological modes diminishes, as
does the ``strength" of standard quantum mechanics, $\hbar$!
The time-dependence of the string as it approaches the ultraviolet fixed
point is reflected in a computation of the string position-momentum uncertainty
relation in the $\sigma$-model deformed by propagating low-lying string modes
and ``instantons". The result for the position-momentum uncertainty, defined
appropriately to incorporate curved gravitational backgrounds, is a
generalization
of the standard uncertainty principle (\ref{8p}), and can be expressed as
\cite{2,3}, for large enough $t$
\[
(\Delta X\cdot \Delta P)_{\rm min}\equiv \hbar_{\rm eff}(t)=\hbar
\left[1+{\cal O}(1/k(t))\right]\qquad\qquad(8')
\]
 The case of the
fundamental length $l_{\rm string}$ is similar. Close to the topological
phase, the
string tension $T\to0$, or $\alpha'\to\infty$, or $l_{\rm
string}(\sim\sqrt{\alpha'})\to\infty$,
while away from the topological phase $\alpha'\to{\rm finite}$, thus $l_{\rm
string}$
gets smaller as time ``runs". Furthermore, since the {\em fundamental} string
length,
 $l_{\rm string}$ contributes an {\em extra term} to Heisenberg's
uncertainty
principle, and since close to the topological phase the uncertainty principle
works at ``full strength", we naturally expect $l_{\rm string}\to\infty$ as
$t\to0$,
while $l_{\rm string}\to{\rm finite}$ as $t\to\infty$, once more a {\em
monotonic
decrease} with time. Indeed, we get \cite{2,3} a generalization of (\ref{20})
that reads, for large enough $t$
\[
l_s(t)=l_s\left[1+{\cal O}(1/k(t))\right]\qquad\qquad(20')
\]
Concerning the cosmological constant $\Lambda_C$, keep
in mind that, as (\ref{31}) indicates, $\Lambda_C$ is proportional to $\delta
c$,
while through (\ref{35}),(\ref{36}), $\delta c=c_m-25=12Q^2$, which is
{\em monotonically decreasing} \cite{2,3} with time, towards its (critical
string)
vacuum
value $\delta c=c_m-25=0$, thanks to the irreversible RG trajectory flow,
alias
Zamolodchikov's $C$-theorem (\ref{52})!

It should be stressed that the above properties are rather general, as based
on
first principles. Indeed, Zamolodchikov's $C$-theorem (\ref{52}) {\em entails}
an irreversible decrease with time of the $C$-function, whose dependence on
the $k$-parameters for large $k$ is of the form $C_0+{\cal O}(1/k)$, for
physically {\em
relevant} CFTs.
Thus we get a monotonic increase with time of $k$-parameters! It looks like
based
upon very general and powerful first principles of string theory, the
``constants"
of Nature, $c,\hbar,l_{\rm string},\Lambda_C$, {\em irreversibly} and
{\em monotonically} decrease with time \cite{2,3}. It looks very plausible that
the
Master
Equation (\ref{60}) is nothing else but the long-sought {\bf N}-equation
(\ref{16})!

Having discussed at length {\em density matrix mechanics}, the ``answer" of
non-critical string theory to the fundamental question of how quantum physics
and gravity best mix, we need to move on to provide ``answers" to the other
``small" problems discussed in section 2 and elaborated further in section 3.

\subsection{Spontaneous collapse of the ``wavefunction" -- classical from
quantum -- the Aphrodite Mechanism}
It must have been pretty clear by now that we have in our hands all the right
ammunition to attack the problems related to the {\bf R} part of quantum
mechanics (see (\ref{15})) such as the Schr\"odinger's cat ``paradox", the
EPR ``paradox" and more generally the emergence of the classical world
from the quantum world {\em dynamically} and {\em spontaneously}
as it should be (see section 3). The spontaneous ``{\em openness}" of the
low-energy world, due to the spontaneous truncation of the topological,
extended modes, coupled to the low-energy propagating particles (the
{\bf P}{\em rocrustean principle} discussed in section 6), is the long-sought
``panacea"!

Before discussing the r\^ole
of topologically non-trivial
configurations on the world-sheet in
the suppression of coherence at large times,
we review a similar phenomenon
in Hall conductors, namely the
suppression of spatial correlations by
``de-phasons'' \cite{39}. A black-hole model is analogous
to a fractional Hall conductor \cite{40},
with the Wess-Zumino level parameter $k$ corresponding
to the transverse conductivity.
Hall systems generally are described
by appropriate $\sigma$-models with Wess-Zumino $\theta$-terms,
defined on the two-dimensional space of electron motion \cite{39}. The fields
of such
$\sigma$-models, which are space-time coordinates in the
black-hole case, correspond to electrons
propagating in the plane,
with the transverse and longitudinal
conductivities $\sigma _{\mu\nu}$
corresponding to background fields in the black-hole case.
The Wess-Zumino terms are associated with instantons
that renormalize non-perturbatively these conductivities \cite{41}:
\begin{equation}
   \beta _{\mu\nu}=\frac{d\sigma _{\mu\nu}}{d \ln L}\ne 0
\label{64}
\end{equation}
where $ L $ is an infrared  cut-off on the instanton size
that serves as a renormalization group scale \cite{41}.

We believe that localization in Hall systems
is directly related to our problem of quantum coherence.
In the {\bf I}nteger {\bf Q}uantum {\bf H}all {\bf E}ffect
(IQHE) model \cite{39}, impurities
are responsible for the localization
of the electron wave function in the plane.
The localization is achieved formally
by representing collectively the effects of impurities
on electron trajectories via extended, static
scattering centres termed ``de-phasons'',
which trap the
electron waves into localized
states with sizes $O(1/\sqrt{\rho})$,
where $\rho$ is the de-phason density.
As a result, the electron correlation functions
are suppressed at large spatial separations:
\begin{equation}
   \propto exp[-(x-y)^2 \rho]
\label{65}
\end{equation}
at zero magnetic field ($\theta = 0$).
As the magnetic field is varied so that
the transverse conductivity becomes a  half-integer
(in units of $e^2/h$),
corresponding to a discrete value of the
instanton angle  $\theta = \pi$, the property
of the
de-phasons to destroy phase coherence between
the advanced and the retarded electron propagators
is lost. Quantitatively \cite{39}, the expectation value of an
electron loop that encircles a de-phason,
in the presence of a magnetic field, is
proportional to
$e^{-(x-y)^2\rho cos\frac{\theta}{2}}$.
Thus, for $\theta = \pi$ the
``effective de-phason density '' $\rho cos(\theta /2)$
vanishes, and
the electrons delocalize implying
a non-zero longitudinal
conductivity.
This delocalization property
is responsible for the transition
between two adjacent plateaux of the
transverse conductivity in the Hall
conductivity diagram \cite{39}.
These ideas can be
extended to the Fractional QHE \cite{42} via the
three-dimensional anyonic Chern-Simons
theories, which are closer
to our black-hole interests. In fact,
it appears to be the zero-field Hall
effect that describes physics at the space-time singularity. For example,
the massive topological modes of the $SL(2,R)/U(1)$
black-hole model (see section 8) are the analogues of the
de-phasons.
As discussed earlier in this section, the ``instantons"
 renormalize the Wess-Zumino
level-parameter $k$ (c.f. $\theta$),
changing the mass and size of the black hole.
The delocalized phase at $\theta =\pi$
may be identified with the ``topological''
phase
at the space-time singularity \cite{36}, which is an infrared
fixed point. The propagating
``tachyon''  mixes in this limit, as
we have discussed above, with the
delocalized topological modes of the string that are
analogous to the de-phasons.
The localization properties are consistent
with shrinking of the world-sheet as
one approaches the ultraviolet fixed point that
corresponds to a flat target space-time where the
tachyons are normal localized fields that do not mix with
topological modes.

Our formalism for \cite{2,3,30} the
time evolution of the density matrix is analogous
to the Drude model of quantum
friction \cite{43}, with the
massive string modes playing the r\^oles of
`environmental oscillators'. In the language of
world-sheet $\sigma$-model couplings $\{ g \}$,
the reduced density matrix of the observable states is given, relative
to that evaluated in conventional Schr\"odinger quantum mechanics,
by an expression of the form
\begin{equation}
 \rho (g, g',  t  ) / \rho_S (g, g',  t  )
\simeq
e^{-\eta \int_0^{t} d\tau \int_{\tau ' \simeq \tau }
d\tau '
 \beta ^i G_{ij} \beta ^j } \simeq
e^{ - Dt ({\bf g}  - {\bf  g'} )^2 + \dots }
\label{66}
\end{equation}
where $\eta$ is a calculable proportionality coefficient,
and $G_{ij}$ is the Zamolodchikov metric (\ref{50})
in the space
of couplings. In string theory, the identification of the target-space
action with  the Zamolodchikov $C$-function $C(\{g \})$ \cite{23}
enables the Drude exponent to be written in the form (\ref{52})
$-\beta^iG_{ij}\beta^j = \partial _t C(\{g\})$,
which also determines the rate of increase
of entropy (\ref{63}) ${\dot S}=\beta^i G_{ij}\beta^j S$.
In the string analogue  of the Drude model
(\ref{66}) the r\^ole of the coordinates in (real) space
is played by the $\sigma$-model couplings $g^i$ that are
target-space background fields. Relevant for us
is the tachyon field $T(X)$, leading us to interpret
$(g-g')^2$ in (\ref{66}) as \cite{30}
\begin{equation}
(g-g')^2=(T-T')^2 \simeq (\nabla T)^2 (X-X')^2
\label{67}
\end{equation}
for small target separations $(X,X')$. Equation (\ref{67})
substituted into (\ref{66}) gives us a
suppression very similar to the IQHE case (\ref{65}).

The effect of the time-dependences (\ref{66},\ref{67}) is
to suppress off-diagonal elements
in the target configuration space
representation of the out-state
density matrix :
\begin{equation}
\rho _{out} (x,x') ={\hat \rho }(x)\delta (x-x')
\label{68}
\end{equation}
This behaviour can be understood intuitively
as \cite{30} being related to the apparent shrinking of the string
world sheet in target space, which destroys interferences
between strings localized at different points in target
configuration space, c.f.
the de-phasons in the Hall model \cite{39}. This
behaviour is generic for
string contributions to the space-time foam, which
make the theory supercritical locally, inducing renormalization
group (target time) flow. Calculations in simplistic models \cite{30}, as well
as dimensional analysis (see in particular the discussion below (\ref{60})),
indicate that the {\bf D}umping (or {\bf D}rude) coefficient $D$ in (\ref{66})
is given by \cite{44,30} $D\approx N\left({m^6\over M^3_{Pl}}\right)$, where
$N$ is the
number of microconstituents of mass $m$. Inserting this value of $D$ in
(\ref{66}), and taking into account (\ref{67}), we arrive at a spontaneous
collapse of the system ``wave-function" with the following characteristics:
localization of the
center-of-mass coordinate $X_{\rm C.M.}$ within the Bohr radius $a_B$
($\sim0.53\AA$) within ${\cal O}(10^{-7}{\rm sec})$,  {\em iff} $N(\approx
M_{Pl}/m)\approx {\cal O}(N_{\rm Avogadro})$!
A rather remarkable result. For us \cite{44,30}, $N_{\rm Avog}\approx10^{24}$
is an {\em indicative} number of constituents for systems showing
characteristics
of ``classical" behavior! What we find here is exactly that: we get a
spontaneous, {\em quick} collapse of the wave-function, for $N\approx
M_{Pl}/m$, not very different from $N_{\rm Avog}$. It is incredible that the
{\em smallness} of Newton's gravitational constant, with respect to the other
interaction coupling constants, or equivalently, the {\em vast difference}
between the low-energy masses (GeV-TeV region) and the Planck mass
($10^{19}\GeV$), is responsible for the {\em almost exact} quantum mechanical
behavior of {\em microsystems} ($N\ll<1$), electrons, quarks, ..., while it
demolishes in no time the wavefunction of {\em macrosystems} ($N\sim N_{\rm
Avog}$). The Schr\"odinger's cat is either dead or not, as the case may be
\cite{44,30},
but not half-dead, half-alive, anymore. There is no Schr\"odinger's cat in
density matrix mechanics. Clearly, such a dynamical mechanism, triggering the
{\em spontaneous collapse} of the wavefunction, will have far-reaching
consequences for the ``measurement" problem in Quantum Mechanics.
 I cannot resist of making a side comment here. The
{\em vast disparity} between $m_{\rm (light)}$ and $M_{Pl}$, known as the
notorious {\em gauge hierarchy} problem in particle physics, has been the
main motivation for practical uses of supersymmetry and supergravity. Actually,
in {\em no-scale supergravity} \cite{45}, one gets {\em naturally} and {\em
dynamically}
a relation of the form $m_{\rm light}\approx e^{-1/\alpha}M_{Pl}$, with
$\alpha$ some characteristic fine-structure-like constant; while the
cosmological constant $\Lambda_C$ is {\em at most} of ${\cal O}(M^4_W)$,
even after the spontaneous breaking of supersymmetry. Recalling also the
fact that the {\em no-scale supergravity framework} seems to be the infrared
limit of string theory \cite{46}, corresponding to  flat-directions
($\beta_g=0$ on
a continous, critical line, not at just a point!), it looks like the {\em gauge
hierarchy problem}, was {\em a blessing in disguise}! In fact, since the
spontaneous collapse of the wavefunction in non-critical string theory is
a well tabulated dynamical mechanism, it leads to some, as it should, specific
experimental predictions. Namely, the exponential decay with time of the
electronic currents in a SQUID with a Josephson Junction at very low
temperatures \cite{44,30}, as well as CPT-violating effects in the $K^0-\bar
K^0$ system \cite{7,47,48,49}. While CPT-invariance is a {\em basic} theorem of
point-like QFT, this is {\em not necessarily}
the case in non-critical string theory. The CPT-theorem is based on {\em
locality}, {\em Lorentz invariance}, and {\em unitarity}. The first two
assumptions are not automatically satisfied in non-critical string theory,
thus it is appropriate to re-open the possibility of CPT-violation.
It goes far beyond my purposes here to discuss these very interesting
proposals, since they have been discussed very clearly and in full detail
in the recent literature \cite{47,48,49}. It suffices to say that these
experiments are
under the way to completion and we may know something not it the very far
future.

Concerning the EPR ``paradox", the very existence of the {\em delocalized},
topological, extended states in continuous interaction with the low-energy
world particle states, turns the low-energy world into an {\em effective
nonlocal theory}, in accord with Aspects's experimental results \cite{10}. In a
way we are dealing {\em effectively with a nonlocal hidden variable theory}
provided by the non-trivial mixing of the low-energy particle modes to the
massive, topological modes induced by the {\em infinite symmetry} content
of non-critical string theory. In an intuitive way, as the two photons,
decay products of the $\pi^0$ (see (\ref{11})), run away in spacetime, still
they are entangled through the topological extended states, and thus when
``something" happens to one photon, say $\gamma_1$, the other ``feels" the
{\em influence instantaneously} due exactly to the {\em very nature} of the
delocalized, topological states. Furthermore, because the very existence of
the topological states is the {\em main reason} for the {\em arrow} of time,
through the {\bf P}rocrustean principle, and thus the {\em un-relativistic}
behavior of the ${\bf R}$-part of quantum mechanics (\ref{15}), clearly the
Penrose ``paradox" gets automatically resolved! If the two ``observers"
wish, they may ``log on" to the indisputable irreversible time ``flow", that
characterizes
the {\bf R}-part of quantum mechanics, and thus straighten out their
differences, {\em getting a universal picture of physical reality}, as it
should be!

As has been repeatedly emphasized above, the dynamical origin of the
spontaneous
collapse of the wavefunction provides a very appealing resolution to the
``classical" versus ``quantum" world problem. The ``{\em coherent phase}"
{\em erosion factor} (second RHS of (\ref{60})) depends {\em explicitly}
on the mass (maybe more precisely on the energy) of the system under
consideration, and while it is {\em virtually harmless} for microsystems
(\eg, electrons, photons, ...) it becomes {\em dominant} for macrosystems
($N_{\rm const}\sim N_{\rm Avog}$). Thus the natural emergence of the
``classical" world from the ``quantum" world. It is amazing, how lucky
the founding fathers of quantum mechanics were. How lucky? About
$(M_{Pl}/m_{\rm
light})^n$, $n=1,2,...$, in other words {\em very lucky}!

Very recently (see second Ref. in \cite{3}), we have been able to show the
emergence of {\em almost-time-reversible} local field theory structures,
associated with decoherence-induced pointer states in the coupling constant
space, as a result of the inevitable couplings to the unobservable
non-propagating solitonic string states. Thus, we satisfy successfully the
last {\em desideratum} of section 3, concerning the {\em dynamical}
appearance, at large distances, of an {\em almost} relativistic QFT.

This is our answer \cite{2,3,25} to the often asked question: why does quantum
mechanics
seem to work so {\em extremely} well? It is just a matter of scale! Since
the {\em de-phasing, erosion mechanism} in target-space is due to the
existence of spacetime foam, thus implying the emergence of the ``classical"
world from the foam, like another Aphrodite, it is appropriate to call this
mechanism the {\bf A}{\em phrodite mechanism}, an immediate consequence of the
{\bf P}{\em rocrustean principle}. It seems that the ancient Greeks had their
mythology straight!

\subsection{A new theory of black-hole dynamics -- No-Horizon Cosmology}
The emerging new picture of black-hole dynamics in non-critical string theory
is not hard to spell. Black holes correspond \cite{28} to topological defects
on the
world-sheet, \eg, monopole-antimonopole pairs, where the monopole, say
placed at the origin of the complex plane (or the south pole of the Riemann
sphere) corresponds to the black-hole (event) horizon, while the antimonopole
placed at infinity (or the North pole of the Riemann sphere) corresponds to
the region of the {\em would be} classical space-singularity. The existence
\cite{28}
of topological defects on the world-sheet presume the non-trivial enlarging
of the world-sheet symmetries, \eg, the existence of $W_{1+\infty}$ symmetries,
that through the EGO mechanism \cite{37} are elevated to target-space
(spontaneously broken) local symmetries. Thus, the stringy black holes are
endorsed {\em naturally} with infinite {\em quantum hair}, \eg, ``$W$-hair"
\cite{33,1},
of the Aharonov-Bohm (AB) type discussed in section 3, thus enabling them to
absorb {\em coherently} vast amounts of information \cite{33,1}.

Any type of interaction/perturbation with/of the black hole will bring it to
some different level, vis-\`a-vis the infinity of AB ``charges" that are needed
for the complete characterization of the black hole identity. The black hole
decay, due to quantum effects \cite{31}, is also an orderly process,
characterized by
a vast set of selection rules \cite{38}, and thus there is no information loss!
Indeed,
string black holes, like massive string modes, are unstable under quantum
corrections. Regularization of divergent integrals over large tori, related
to modular invariance, {\em a purely stringy effect}, induce a non-zero
${\em Im}(M_{\rm BH})\propto 1/\tau_{\rm BH}$, \ie, causing the decay of the
black hole \cite{31}. An explicit calculation shows that the 4-D black-hole
lifetime \cite{31,50}
is given by $\tau_{\rm BH}\propto M^3$, very similar to Hawking's result
(\ref{7})! Of course, in our case we have a {\em normal, ordinary, coherent}
decay process, like $\Delta^{++}\to p\pi^+$, and not Hawking's, thermal-like
radiation. The formal proof \cite{33,1} of these statements rely on the fact
that, \eg,
the $W_{1+\infty}$ symmetry on the world-sheet is just an area-preserving
diffeomorphism on the phase-space of the ``coupling constants", in other
words, imply vanishing of $\delta\H$ (see (\ref{60})), thus no information
loss! On the other hand, we should be able \cite{38} to ``measure" all the AB
``charges",
as explained in section 3, in order to get the {\em complete} picture. Since
this is practically impossible, the {\em Procrustean principle} intervenes and
{\em effectively} ``opens" the black hole system, turning it into a NEQSS!
Actually, we have proven explicitly \cite{50} that when we ``sum over" the
unobservable
AB ``charges", \eg, ``$W$-hair", we are getting back the standard
Beckenstein-Hawking formula (\ref{6}). The proof goes like this \cite{50}:
stringy
black holes, like the massive string modes, are characterized by a
level-multiplicity
\begin{equation}
N(M_{\rm BH}){\buildrel M_{\rm BH}\to\infty\over\approx}
e^{2\pi\sqrt{\alpha'}}e^{\sqrt{2+Q^2}M}\approx e^{2\pi\alpha'M^2_{\rm BH}}
\label{69}
\end{equation}
thus implying a 4-D black-hole entropy (by ``summing over" the AB ``charges")
\begin{equation}
S_{\rm BH}\sim\ln N(M_{\rm BH})\sim M^2_{\rm BH},
\label{70}
\end{equation}
which is nothing else but (\ref{6})! Of course, here we have an explicit
{\em microscopic mechanism} that explains how starting from a {\em consistent}
and {\em complete} quantum theory of gravity, we end up with some {\em apparent
loss of information}, carried away by the topological modes, due to our
``confinement" in a local, low-energy world! In a way, it looks like the
well-known {\em energy crisis} or the {\em energy catastrophe} of 1929, that
led eventually to the discovery of the neutrino. A {\em continuous} spectrum
as a function of the electron energy, implies that either $\beta$-decay is
a two-body process, in which case  {\em energy} is {\em not conserved}, \ie,
the
{\em energy crisis} or {\em energy catastrophe} supported strongly by Bohr;
or it is a three-body process involving a new particle ({\em neutrino}) that
entails the partioning of the available energy between the neutrino and the
electron, resulting in a continuous electron energy spectrum, as suggested by
Pauli. Bohr was wrong!  The so-called
{\em Hawking catastrophe} (discussed in section 2), as far as non-critical
string theory is concerned, descends to the same level as the {\em energy
catastrophe} above, where the role of the {\em elusive neutrino} is played
by the {\em delocalized, topological states}.
While the thermodynamic laws of black holes {\em are
derivable} in non-critical string theory, they are just that: thermodynamic
laws,
\ie, a coarse-grain approximation, neglecting the nitty-gritty details of the
involved topological modes. Actually, this new picture \cite{1,2,3} provides a
crytal-clear answer to the frequently asked question: what is the fate of the
black hole? The answer is: it disappears into the spacetime foam! Indeed, as
the black hole decays {\em coherently}, it loses mass and eventually it
becomes indistinguishable \cite{28} from the virtual black holes that are part
of the spacetime foam, having transmitted all its information to its decay
products, which include the topological modes \cite{28}. Another frequently
asked question is:
what happens to the physically intuitive Hawking mechanism for the evaporation
of the black hole, where near the BH horizon a particle-antiparticle pair is
spontaneously generated and then one falls into the BH, while the other arrives
at the ``observer" at infinity, who thinks that the BH radiates? The answer is
again
simple: stringy black holes, because of the vast amount of ``charges" they
carry, are {\em extreme black holes}, like the usual {\em extreme
Reissner-Nordstr\"om} black holes whose main characteristic is the absence of
Hawking radiation. Thus their ``horizons" are not an
{\em attractive} place to be (the gravitational attraction is balanced by the
``charges" repulsion), and no member of the particle-antiparticle pair is able
to fall into the black hole, and thus the particle-antiparticle pair is
spontaneously annihilated, and the ``observer" at infinity sees nothing!

Concerning the black hole ``classical" space-singularity, the new stringy
black-hole theory provides a very satisfactory and aesthetically appealing
answer: there is no space-singularity! Let me explain. As we discussed in
the previous section, as we approach the would-be classical ``singularity",
spacetime dissolves ($\VEV{g_{\mu\nu}}=0$) and the standard Riemannian-like
notions of spacetime break down, leaving topological field theory as the
only suitable description of this classical ``singular" region \cite{36}. In a
way, the region of infinitely many symmmetries cannot support a spacetime
interpretation, it suffers a phase transition characterized by the vanishing
of $\VEV{g_{\mu\nu}}$, and thus becomes topological. It looks like the new
theory of
black-hole dynamics has the potential of providing satisfactory answers to all
the standard problems of black-hole physics. Furthermore, black-hole dynamics
are intimately related to the questions concerning the ``{\em initial
singularity}" in the early Universe, and thus we shouldn't be surprised that
we may get satisfactory answers to cosmological problems, as well. After all,
in many cases, and in non-critical string theory in particular, cosmological
solutions (\eg, (\ref{29})) are related to black-hole solutions by a
Minkowski rotation ($r\lra t$) (see section 8). Indeed, very near the
would-be classical
{\em initial singularity}, spacetime dissolves ($\VEV{g_{\mu\nu}}=0$)
and thus the spacetime description is replaced by some topological field
theory characterized by infinitely many symmetries. This topological field
theory corresponds to the {\em eternal topological phase}, that is {\em always
there}, since there is no notion of time in this phase, let it be! Dynamical,
energetically favorable, spontaneous breakdown \cite{36} of some of the
infinitely many symmetries in the topological phase, leads to an {\em expanding
spacetime} ($\VEV{g_{\mu\nu}}\not=0$) Universe \cite{36}. {\em Why spacetime}?
Because the metric
tensor (together with many other fields) gets a non-zero vacuum expectation
value, thus allowing a ``standard" description of our Universe.
{\em Why expanding}? The topological phase, or the region very-very close
to it, is characterized by a very high value of ``central charge deficit"
$\delta c=c_m-25=12Q^2$ (see (\ref{30}),(\ref{35}),(\ref{36})) ``rolling"
down irreversibly along the RG trajectory, according to Zamolodchikov's
$C$-theorem \cite{23} (\ref{52}). This fact, combined with our interpretation
of the
RG scale as time (\ref{49}), makes $Q^2$ a monotonically decreasing function
of $t$, and thus through (\ref{32}) leads to an initially fast
($Q^2(t\to0)\gg1$) and much later slow ($Q^2(t\to\infty)\to0$) {\em expanding}
Universe. A rather dynamical and natural explanation \cite{2,3} of the origin
of the
expansion of the Universe. Furthermore, very close to the topological phase
($t\to0$), as discussed above, the velocity of light $c\to\infty$ (not
unrelated to the fact that $\delta c(t\to0)\to\infty$) implying that the
horizon distance $d$ in co-moving coordinates over which an observer can
look back $d=\int c(t)dt\to\infty$, or much larger than the naive estimate
$d=ct=({\rm constant})t$. In other words, as $t\to0$, {\em all} regions were
{\em communicado}, and {\em not} as naive extrapolations of the standard
Big-Bang cosmology suggest that there were, say at $t\approx10^{-35}$sec
about $10^{80}$ {\em incommunicado} regions! This is the simplest, to me,
solution \cite{2,3} of the {\em horizon problem}. As, $t\to0$, there are {\em
no horizons}, because $c\to\infty$ and thus the light-cone gets {\em
flat-open}.
Actually, the similarity with the EPR paradox is striking: (a) the {\em
no-horizon} embryonic Universe {\em makes it possible} to define initially
a ``coherent" wavefunction $\Psi_{\rm Universe}$, corresponding to the
$\Psi_{\rm system}$ (\ref{11}) of the two photons, implying that intially
all the regions in the Universe were in ``contact" and in entanglement,
while (b) later on ``measurements" on one local region led to a controllable
disentaglement through the {\em influences} of the topological modes, that
trigger any other ``disconnected" local region to (quantum) ``jump" on a
specific,
strongly inter-related state, corresponding to the ``quantum jump" that
photon $\gamma_2$ suffers, after ``measuring" photon $\gamma_1$. The presently
observed smoothness, \ie, homogeneity and isotropy of our Universe, as measured
by the variation of the cosmic background radiation ($\Delta
T_0/T_0\approx10^{-5}$), over angular scales covering effectively the whole
sky, should be as surprising as the two-photon correlations in Aspect's
experiment \cite{10}! It is just
{\em nonlocality}, a fact of quantum mechanical life, that finds a very
interesting microscopic explanation through the existence of the topological
modes, \ie, the Procrustean principle, in non-critical string theory. It
should be stressed that in the {\em non-critical} Universe, the origin of
horizons, \ie, light-cones, is {\em dynamic} as occuring during the phase
transition from the topological phase ($c\to\infty$) to the spacetime phase
($c\to{\rm finite}$). An appropriate name would be {\bf NO}-{\bf HOR
izon cosmology} implying {\em not the absence of horizons}, but {\em the
existence of dynamically created horizons}!\footnote{Any similarity with
the {\em no-scale supergravity framework} \cite{45}, implying {\em not the
absence of (mass) scales}, but {\em the existence of dynamically created (mass)
scales}, is {\em not accidental}!} Furthermore, if the dependence on $t$ of the
velocity of light $c=c(t)$ is a smooth function (as it is for $t\gg0$) we can
always redefine \cite{2,3} it to be 1, thus avoiding any conflicts
with special relativity. The variation of $c$ with cosmic time $t$ should
have some experimental consequences in phenomena that evolve strongly with
cosmic time, and are due to processes in the very early, fast expanding,
violent Universe. We have already mentioned one, \ie, the absence of horizons
at the very beginning ($t\to0$) due to the very fact that as $t\to0$, $c=c(t)$
becomes a singular function, and cannot be reset to 1! Studies of other
phenomena are in progress. Concerning the problem of the presently large
entropy content of our Universe, {\em no-horizon cosmology} provides a very
simple answer \cite{2,3}. It is the entropy that has been generated according
to
(\ref{63}) while the {\em non-critical Universe} ($\beta^i\not=0$) ``rolls
down" its RG trajectory, starting at the topological phase (infrared fixed
point) and {\em eventually} arriving at the critical string vacuum (ultraviolet
fixed point). We should be very pleased that we are not yet at the UV-fixed
point, because that would mean the End of Time! Clearly, by having turned our
``low-energy {\em observable} world" into an ``open" NEQSS, the
increase in entropy, as irreversible time ``flows", should be no paradox.
A very different situation, indeed, from the {\em quite} adiabatically
({\em isoentropically}) expanding standard Big-Bang Universe! Of course, at
very late times, presumably like the ones we are now living, the rate of
increase of entropy is reduced significantly ($\beta^i\to0$), as (\ref{63})
indicates, thus implying asymptotically a standard Big-Bang adiabatically
expanding Universe. In a way, all the information that the delocalized
topological modes have carried away, since the {\em beginning of time}, is
quantitatively expressed by the presently large entropy content of our
Universe. It should be stressed that the generation of entropy in the
non-critical Universe is only natural since we are dealing with an ``open"
system. Futhermore, the rate of increase of entropy (\ref{63}) is
tremendously enhanced very close to the topological phase ($Q^2(t\to0)\gg1$)
since there the topological modes are very ``active", carrying away a lot of
information, thus leading to a fast increase of entropy. This huge production
of entropy at the very first stages of the non-critical Universe, just out
of the Topological Phase, has some very desirable, long-sought features.
It leads to the natural dilution of all kinds of problematic, undesirable,
either topological defects (\eg, monopoles, domains, etc.) or particles
(gravitini, hidden sector ``cryptons", etc.), thus helping to provide
swiftly an {\em enviromentally clean}, smooth Universe. In addition, it
certainly helps to diminish dynamically the {\em space-curvature} $k_s$.

The {\em space curvature} $k_s$ in the non-critical Universe is either
{\em zero},
as (\ref{33}) implies for any $D$ (including $D=4$), {\em or} in the particular
case
of $D=4$ there is another possibility available, $k_s\propto 1/k$, with $k$
the
level parameter of the world-sheet Wess-Zumino (WZ) model on the group
manifold SO(3) (see discussion after (\ref{30})). If we insist on a WZ type
of
solution, since the level parameter $k\to\infty$ for large ($\gg1$ in stringy
or
Planck units) space-times, it follows that $k_s(\propto1/k)\to0$. Thus, in
either
case ((\ref{33}) or WZ type of solution) the net outcome is that for large
space-times \cite{18}
\begin{equation}
(k_s)_{l\gg l_{\rm string}}=0
\label{71}
\end{equation}
\ie, a {\em spatially-flat} Universe! This is the dynamical, stringy,
microscopic
resolution of the {\em flatness problem}, as provided by the {\bf NO-HORIZON
COSMOLOGY}, in full accord with the huge production of entropy at the
initial stages of the non-critical Universe, as discussed above.

Concerning the cosmological constant $\Lambda_C$, we have argued (second
Ref. of \cite{33}) that the infinite set of string symmetries that preserve
quantum coherence may also be responsible for the vanishing of the cosmological
constant. If this is indeed true, then we expect that, as in the case of
quantum coherence, the Procrustean Principle intervenes and renders the
cosmological constant {\em effectively} non-zero, and ``{\em running}".
Indeed, its dynamical evolution
with time (for any $D$) is given by \cite{2,3}
\begin{equation}
\Lambda(t)={\Lambda(0)\over 1+t\left({\Lambda(0)\over  D-1}\right)}
\label{72}
\end{equation}
with $\Lambda(0)=2/\alpha'$, implying an asymptotically-free cosmological
constant $\beta$-function, thereby leading {\em dynamically} and {\em fastly}
to a vanishing
cosmological constant at the ultraviolet fixed point on the world-sheet,
corresponding to the critical string vacuum, $\delta c=0$. This is the
microscopic realization of the quantum relaxation mechanism \cite{2,3} for the
vanishing of the cosmological constant, advocated at the beginning of
this section. While it is premature to discuss quantitatively the implications
of this specific mechanism for the vanishing of $\Lambda_C$, it is
amusing to point a few relevant points. The present age of the Universe
 is about $10^{60}$ in natural (Planck) units. Another
point is that one-loop calculations at the point-like field theory level in
{\em no-scale supergravity models} \cite{45}, discussed above in conjunction
with the resolution of the Schr\"odinger's cat problem, yield a negative
contribution to the cosmological constant that is ${\cal O}((M_W/M_{Pl})^4)
\sim10^{-60}$ in Planck units. Finally, we note that astrophysical and
cosmological observations are {\em compatible} with a present-day value
of about $10^{-120}$ in Planck units. Moreover, a cosmological constant
of this order of magnitude could even be a welcome adjunct to Cold Dark
Matter models. Thus it may be even {\em desirable} that the cosmological
constant $\Lambda_C$ has not yet completely relaxed \cite{2,3,51}! It is highly
remarkable that
{\bf NO-HORIZON COSMOLOGY} in conjunction with the no-scale framework
\cite{45},
both crucial parts of non-critical string theory, may eventually lead through
(\ref{72}), not only to a natural, dynamical resolution of the cosmological
constant problem, but also to some observable consequences as well!

In a few words, the Universe according to \cite{2,3} non-critical string theory
is a
very dynamical NEQSS, where the very origin of space-time, the subsequent
expansion, the huge entropy generation, the large-scale smoothness, and
the apparent vanishing of the space-curvature $k_s$ and of the
cosmological constant $\Lambda_C$ are {\em all} due to a {\em universal
quantum relaxation mechanism}: irreversible ``flow" along the RG trajectory,
starting at the infrared fixed point (eternal topological phase) and ending up
at
the ultraviolet fixed point (critical string vacuum). Once more, I would like
to
stress
the fact that critical string vacua are {\em derivable} dynamically in
non-critical
string theory, and that the apparent similarity of our Universe to a
critical-string
vacuum asymptotically is only due to the diminishingly small
$\beta$-functions,
$\beta_i\approx{\cal O}((E/M_{Pl})^n)$, $n=1,2,\ldots$, at large times.
Actually,
the Hubble expansion parameter, in the {\em no-horizon cosmology}, owes its
origin
to the non-vanishing of the $\beta_i$-functions, thus one very naively may
expect
$H\propto \beta_i$, \eg, $H\propto E^2/M_{Pl}\approx T^2/M_{Pl}$ at large
times, providing a possible microscopic explanation of a well-known result
in Standard Big-Bang cosmology! In addition, the apparently small ``anomalous
dimensions" of the ``matter" deformations may eventually lead in target-space
to {\em quasi-scale-invariant} energy density perturbations $\delta\rho/\rho$,
responsible eventually for galaxy formation and other non-trivial cosmic
structures observed in our Universe. The NEQSS nature of the non-critical
Universe provides a new angle for looking at the cosmic texture problem.

\section{Interlude (III) -- Passatempo}

Until now I have tried to present the physics of non-critical string theory as
generally
as possible, and willingly and consciously I have avoided the use of explicit
models.
As an ``existence" proof, together with John Ellis and Nick Mavromatos, we
have
``derived" explicitly most of the physics I presented in the previous sections
in the
case of a 2-D ({\em space-time}) stringy black hole, due to E.~Witten
\cite{26}. This
``toy"
laboratory shouldn't be taken lightly, as it may take us a long way towards
understanding
how non-critical string theory applies to the real world. There are many
reasons
supporting this allegation. Two-dimensional stringy black holes may be
considered
as spherically symmetric four-dimensional black holes \cite{50}, which as is
well-known,
are
the studying ground of full-fledged 4-D black hole dynamics. Using the
techniques
of matrix models \cite{52} one may sum up the effects of higher genera and
provide
(quasi)
exact solutions, avoiding the usual traps of perturbation theory. The embedding
of the
2-d world-sheet into space-time is easier to handle and the one-to-one
correspondence
between topological defects on the world-sheet and stringy black holes becomes
apparent \cite{28,40}. Furthermore, one relies heavily, again, on the ``linear
dilaton"
solution
(\ref{29}), which is valid for $c_m=1$, provided that $X^0\leftrightarrow
X^1$!
In addition, the world-sheet CFT corresponds \cite{26} to a {\em gauged}
Wess-Zumino model based on a $SL(2,R)/[U(1)\ {\rm or}\ O(1,1)]$ coset space,
not unrelated to the existence of a ``linear dilaton" non-critical string
vacuum,
as discussed in subsection 4.2. Actually, there is a deep reason for the
frequent
appearance of the gauged WZ type models. Non-compact groups/symmetries on
the world-sheet, leading to very interesting non-trivial backgrounds/physics
in target space, are usually represented by unacceptable non-unitarity
current algebras. Only some very specific cosets of the non-compact groups
turn out to be {\em unitary}, and the {\em natural framework} to study them,
by providing explicit Lagrangians, are the {\em gauged} Wess Zumino models
on the corresponding coset spaces! So, it looks that in whatever we do, if
we want to take into account non-trivial topological configurations in
$D$-dimensional {\em target-space} (\eg, taking into account space-time foam),
we better have on the world-sheet a piece of the CFT corresponding to some
gauge WZ coset space. I do hope that by now I have provided enough evidence
and arguments that the Witten 2-D stringy black hole should be taken very
seriously, specifically in its world-sheet representation, as most of its
properties may be independent of the target-space dimension, modulo trivial
rescalings of the different quantities with $D$, as follow from simple
dimensional analysis. As I have emphasized above, together with
John Ellis and Nick Mavromatos, we have gone through all the ``steps" that
I have presented previously in the two-column format, and we have
tabulated our results elsewhere \cite{1,2,3}. So there is no need here to
repeat them
again, except for emphasizing a few crucial points.

The action of the model is \cite{26}
\begin{equation}
   S_0=\frac{k}{2\pi} \int d^2z [\partial r {\overline \partial } r
- \tanh^2 r \partial t {\overline \partial } t] + \frac{1}{8\pi}
\int d^2 z R^{(2)} \Phi (r)
\label{73}
\end{equation}
where $r$ is a space-like coordinate and $t$ is time-like,
$R^{(2)}$ is the scalar curvature, and $\Phi$ is the dilaton field. The
customary interpretation \cite{26} of (\ref{73}) is as a string model with
$c_m$ = 1 matter, represented by the $t$ field, interacting with a
Liouville mode, represented by the $r$ field, which has $c_m  < 1$ and
is correspondingly space-like. As an illustration of the
approach outlined in the previous sections, however, we re-interpret
\cite{30,2,3}
(\ref{73}) as a fixed point of the renormalization group
flow in the local scale variable $t$. In our interpretation, the
``matter'' sector is defined by the spatial coordinate $r$, and has
central charge $c_m$ = 25 when $k  = 9/4$. Thus the model
(\ref{73}) describes a critical string in a dilaton/graviton
background. The fact that this is static, i.e. independent of $t$,
reflects the fact that one is at a fixed point of the renormalization
group flow.

The Witten 2-D black hole \cite{26} is endorsed with $W$-hair \cite{33,1}, due
to the existence
of a non-trivial $W_{1+\infty}$ algebra, that is, a ``coupling constant"
phase space area-preserving symmetry, thus implying $\delta\H=0$, \ie, no
loss of quantum coherence \cite{33,1}! Nevertheless, Chaudhuri and Lykken have
argued \cite{53} that the exactly-marginal
deformation that turns on a static tachyon background for the Witten
black hole necessarily involves the higher-level topological string
modes, that are non-propagating delocalized states, which are interrelated by
an infinite-dimensional $W$ symmetry. This is a consequence
of the operator product expansion of the tachyon zero-mode operator
${\cal F} _{-\frac{1}{2},0} ^c    $ \cite{53}:
\begin{equation}
    {\cal F} _{-\frac{1}{2},0}^c    \circ
    {\cal F} _{-\frac{1}{2},0}^c    = {\cal F}_{-\frac{1}{2},0}^c
+ W_{-1,0}^{hw} + W_{-1,0}^{lw} + \dots
\label{74}
\end{equation}
where we only exhibit the appropriate holomorphic part for reasons of economy
of space. The $W$ operators and the $\dots$ denote level-one and higher string
states.

Since a ``local" observer cannot take into account this mixing of the
tachyon with the delocalized modes, she truncates them, thus unavoidably
getting an effective loss of coherence. Furthermore, we have {\em explicitly}
identified \cite{28,2,3} the topological defects on the world-sheet as
``monopoles"
and ``instantons", that correspond to black holes in target space and have
worked out their contributions to world-sheet correlation functions using the
respectfull ``valley"-method of non-perturbative physics \cite{30,2,3}. Then,
when these
``valley"-improved world-sheet correlation functions get translated to
target-space correlation functions, they lead immediately to {\em
non-factorizable} $\S$, as we have advocated above \cite{30,2,3}. In the
particular case of
the Witten 2-D black hole \cite{26}, the space-time foam is represented
\cite{28,40} on the
world-sheet by a {\em Quantum Hall fluid}, and thus many of the remarkable
properties of Quantum Hall fluids get transmitted to black hole physics
\cite{40}, as
we have emphasized in section 7.

Concerning the problem of the black hole ``singularity" \cite{23}, it has been
shown that the conformal field theory action close to the
singularity can be rewritten in the form \cite{26,35,36}:
\begin{equation}
S=-\frac{k}{4\pi} \int d^2x D_i a D_i b + i \frac{k}{2\pi}
\int d^2x w \epsilon^{ij}F_{ij}
\label{75}
\end{equation}
where the singularity is parametrized \cite{26} by the limit $uv = 1$ of the
Kruskal-Szekeres coordinates $u$ and $v$ which have been written in
the forms $u = exp(w), v = exp(-w),$ and $a, b$ are diagonal elements in the
general $2 \times  2$ $SL(2,R)$ matrix - the off-diagonal elements being $u$
and $-v$ with $ab + uv =1$. In equation (\ref{75}), $F_{ij}$ is the field
strength of the $U(1)$ gauge potential $A_i$, and the indices $i,j$ take two
values, corresponding to the two dimensions of the space-time, and the two
variables $a$ and $b$ both vanish at the singularity. The exciting and key
observation \cite{26,35,36} is that the residual theory resembles a {\it
topological field
theory}, namely the dimensionally-reduced form of a Chern-Simons theory in
three dimensions.

That a topological field theory should emerge at the singularity was
perhaps not surprising for at least one reason. It is thought that topological
field theories correspond to an unbroken phase of gravity, in which the metric
vanishes, and it is precisely at the singularity that there is no
physically-meaningful metric. Here there is an analogy with the core of
a magnetic monopole, where the vacuum expectation value of the Higgs field
vanishes at the centre of a topologically-non-trivial solution of the
field equations, and the underlying gauge symmetry is restored.  In the same
way, we expect that a non-zero value of the metric is at least one of a
possibly infinite number of order parameters that mark the spontaneous
breakdown of a higher stringy symmetry, which should become manifest in the
associated topological field theory.

As described in previous sections, we have identified \cite{33,1} the essential
symmetry that safeguards quantum coherence as a $W$-symmetry, with
an infinite set of associated conserved charges that are in principle
measurable \cite{38}. Indeed, there is evidence that the topological field
theories
relevant to space-time singularities have a high degree of symmetry that
includes $W$-symmetry!

It should be noticed that in the case of 2-D stringy black holes \cite{26},
their
mass $M_{\rm BH}\propto (k-2)^{-1/2}$ and the velocity of light
$c\propto c_0\sqrt{k\over k-2}$, where $k=k(t)$ is the level parameter of
the Wess-Zumino-Witten model (\ref{73}), thus showing {\em explicitly}
\cite{2,3}
some of our repeatedly stressed points: (1) for large space-times,
corresponding to $k(t\to\infty)\to\infty$, the black-hole mass diminishes,
through quantum decay \cite{31}, and eventually the black hole is absorbed in
the
space-time foam \cite{28}, while the velocity of light $c\to c_0$, its critical
string vacuum, asymptotic value; (2) as $k(t\to0)\to2$, \ie, very close
to the topological phase, the velocity of light $c(t\to0)\to\infty$, due
to the dominance of the topological, delocalized modes and the eventual
breakdown of our naive space-time interpretation of the ``classical"
singularity! Interestingly enough, as (\ref{75}) indicates {\em explicitly},
one basic, fundamental property of gauged Wess Zumino-type models on coset
spaces is their {\em topological nature}, that enables them to metamorphise
into a topological field theory, uppon hitting a would be target space
``classical" singularity.

\section{Back to the Future}
Non-critical string theory seems to provide a suitable framework \cite{1,2,3}
for addressing
all the
``small" problems discussed in section 2. The suggested resolution of these
problems
is characterized by its simplicitly and its universality. Indeed, as we
conjectured in
section 3, we do find that black hole dynamics, the collapse of the
wavefunction, and
quantum gravity are so internally correlated that a satisfactory resolution of
one of
these ``small" problems naturally implies a resolution to the others. We
believe that
the essential brand new element that non-critical string theory brings in is
the
{\em Procrustean Principle}. The issue of {\em nonlocality}, which hovers
around
in conventional Quantum Mechanics, becomes the {\em central issue} in
non-critical
string theory. It is exactly the {\em conflict} between the limited ability of
the ``local"
observer, ``who cannot see beyond his (her) nose", and the existence of {\em
delocalized, topological} states in continuous interaction with the low-energy
string modes, that eventually lead to spontaneous truncation, thus
``opening"
the low-energy observable system ({\em the Procrustean Principle}) with all
its
consequences discussed previously. It is of fundamental importance to
understand
that the possible extension/modification of quantum mechanics and (special)
relativity suggested by non-critical string theory is of a very original and
somehow
subtle nature. If we {\em tacitly assume} that we are living in a {\em fixed}
(Minkowski) {\em space-time}, then {\em critical} string theory may provide a
{\em complete} framework for calculating consistently quantum gravitational
corrections,
but inherits all other ``small" problems of point-like QFT. We claim
\cite{1,2,3} that the
{\em tacit assumption} or {\em premise} of a {\em fixed} (Minkowski) {\em
space-time}
is false, and that this idealistic premise has to be replaced by one that takes
into
account the existence of space-time foam. This possibility exists {\em only}
in
non-critical string theory where space-time foam is naturally taken into
account and
the Minkowski background of critical strings is {\em replaced} by the ``linear
dilaton"
background, similar to (\ref{29}). In such a case, one automatically has to use
the
general approach (\ref{16}), outlined in section 3, which takes a very concise
form \cite{25}
(\ref{60}), as explained in section 7. In the non-critical string theory
framework,
critical string vacua correspond to fixed points of the RG trajectory, while
the big
question of why our Universe looks so much like a Minkowski background
(critical
vacuum) gets a very reasonable answer: the $\beta_i$-functions responsible for
the RG ``flow" are diminishingly small: ${\cal O}((m_{\rm light}/M_{Pl})^n)$,
$n=1,2,\ldots$! In a way, we have extended the two big principles of modern
physics, the quantum principle and the relativity principle to

\begin{center}
{\bf Three Big Principles}
\end{center}

\begin{description}
\item (I) {\em Particle--Wave duality}
\begin{itemize}
\item It leads to the introduction of a new {\em dimensional} constant of
nature,
the Planck constant $\hbar$, with dimensions of {\em action}, diminishingly
small
with respect to the classical world natural units.
\item It leads to the {\em uncertainty principle}
\[ \Delta X\cdot\Delta p\ge\hbar\]
which implies that ``we cannot measure simultaneously the position and momentum
of
a particle", and thus making the quantum world  {\em very different} from the
classical world.
\end{itemize}
\item (II) {\em The laws of nature are {\bf locally} form-invariant under
changing of
inertial frames}
\begin{itemize}
\item It leads to the promotion of the velocity of light $c$ to a universal
dimensional
constant of nature, with $1/c$ diminishingly small with respect to the
classical world
natural units.
\item It leads asymptotically to {\em restricted horizons}
\[ ds^2=c^2dt^2-d\vec x^2\]
which implies that ``we cannot send signals traveling faster than light", and
thus
turning the Newtonian-Galilean notion of absolute ``simultaneity" into a
``relative"
(observer-dependent) one.
\end{itemize}
I conjecture now  that the third ({\em missing}) big principle should read as
follows
\item (III) {\em The laws of nature are effectively {\bf nonlocal} -- there is
no {\bf closed}
system
(a physical extension of Goedel's theorem)}
\begin{itemize}
\item It leads to the promotion of the string fundamental length $l_{\rm
string}$
to a universal dimensional constant of nature, diminishingly small with
respect
to the classical world natural units.
\item It leads to the {\em Procrustean Principle}
\[ g=(g)_{\rm local}\oplus(\sum_i``g_{W_i}")_{\rm nonlocal}\]
with $g$ the (non-critical) string backgrounds, which implies that ``we cannot
measure
it {\em all}", because we are {\em locally} confined, thus we spontaneously
truncate
the system, \ie, we ``open" it, triggering a monotonic increase of entropy, as
recorded
in the apparent  ``{\em arrow}" or ``{\em irreversibility}" of
time.
\end{itemize}
\end{description}

A clarifying remark is {\em badly} needed here. String theory is of course
based on {\em local} quantum field theory on the {\em world-sheet}, and
as far as {\em ``world-sheet" physics} is concerned, the two Big Principles
{\em Quantum} (I) and {\em Relativity} (II) remain {\em exact, intact,
untouchable} according to the {\em old scriptures}! Nevertheless, {\em
locality} on the world-sheet is {\em not equivalent} to {\em locality} in
target spacetime. Thus, the third, Goedelian-like, Big Principle of
(target spacetime) {\em Nonlocality} (III) intervenes and renders {\em
approximate}, as far as {\em ``target spacetime" physics} is concerned, the
{\em Quantum} (I) and {\em Relativity} (II) principles, in sharp disaccord
with the {\em old scriptures}! After all, the scriptures should not be
something just carved in dead stone, but instead we should allow, as past
experience has abundantly taught us, for some revamping and/or addition of
a few {\em new} lines.

It is worth emphasizing that the above three big principles are mainly
concerned
about {\em information}: loss of information (I,III) or transmission of
information (II).
Thus we shouldn't be surprised that NEQSS, as described by non-critical string
theory, play a fundamental role in describing our Universe. Actually, as is
well-known,
in string theory the different constants of nature, $c,\hbar,l_{\rm string}$
are
interrelated, and thus the common origin ($k=k(t)$) of their
quantum-relaxation
towards their asymptotic values, is only natural. Since all the constants of
nature
are functions of the level parameter $k$ characterizing the world-sheet CFT
(\eg, see (\ref{73}),(\ref{75})), it is only natural that all the above big
principles and
their
consequences may be eventually ``derived" from the following

\begin{center}
{\bf Protean Principle}
\end{center}

There is an Eternal Topological phase of our Universe, {\em classically}
perceived
as the {\em initial singularity}, corresponding to a suitable topological field
theory
on the world-sheet characterized by some level parameter(s) $k$, ...

\vspace{2cm}

Well, it is about {\em time} to stop. There is a question that we are unable to
answer
at this {\em time}: does non-critical string theory have anything to do with
the
``works" of the real Universe?\\
Only {\em time} will tell,\\

\begin{center}
but you must remember this,\\
a (standard) model, is just a model,\\
a (point-like QF) theory, is just a theory\\
the fundamental (non-critical) strings apply\\
 as {\em time} goes by ...
\end{center}

\newpage

\section*{Acknowledgements}
It gives me great pleasure to thank my close collaborators John Ellis and
Nick Mavromatos, for a very interesting, thrilling, and adventerous
collaboration, and for reading the manuscript.
 My heartfelt thanks to my friends, David Norton, Vice-President
for Research at HARC, and Nino Zichichi, for their strong encouragement and
support that generously provided to me through all this turmulous {\em time},
for calming me down, as it was badly needed, in some critical moments, and
for their suggestion to write down something {\em simple}. To my close
collaborator and friend, Jorge Lopez, my sincere thanks for being always
there to patiently discuss and provide crucial assessment of my, sometimes
not very coherent outbursts, and for a critical reading of the manuscript.
Many thanks to Giannis Giannakis and Kajia Yuan for a very thorough, critical
reading of the manuscript and for very appropriate and helpful suggestions.
To my greek friends: $\Lambda\epsilon\upsilon\tau\epsilon\rho\eta s$
$\kappa\alpha\iota$ $\Pi\alpha\nu\alpha\gamma\iota\omega\tau\eta s$
$\Delta o\upsilon\kappa\alpha s$ (MDs), my sincere thanks for very interesting
discussions about {\em time}, the {\em collapse of the wave function}, and
the physiology of {\em neurons}. Last, but definitely not least, my wife Myrto
deserves all my gratitude and thanks, from the bottom of my heart, for the
kindness, infinite patience and understanding she has shown for a very long
time now, and in particular, last summer in Santorini where she heard
continously and I am sure more than she cared for, about the nature of {\em
time}!

This work has been partially supported by DOE grant DE-FG05-91-ER-40633.

\end{document}